\documentclass[aps,pra,reprint,superscriptaddress,showpacs]{revtex4-1}

\bibpunct{[}{]}{,}{n}{}{}

\usepackage{graphicx}
\usepackage{amsmath}
\usepackage[caption=false]{subfig}
\usepackage{xcolor}

\begin{document}


\title{Theory, Design, and Experimental Verification of a Reflectionless Bianisotropic \textcolor{black}{Huygens'} Metasurface for Wide-Angle Refraction }


\author{Michael Chen}
\affiliation{{The Edward S. Rogers Sr. Department of Electrical and Computer Engineering,
University of Toronto,\\Toronto, ON, Canada M5S 2E4}}

\author{Elena Abdo-S\'{a}nchez}
\affiliation{{Dpto. Ingenier\'{i}a de Comunicaciones, E.T.S.I. Telecomunicaci\'{o}n, Universidad de M\'{a}laga,\\Andaluc\'{i}a Tech, E-29071 M\'{a}laga, Spain}}

\author{Ariel Epstein}
\affiliation{{Andrew and Erna Viterbi Faculty of Electrical Engineering,
Technion - Israel Institute of Technology, Haifa 3200003, Israel}}

\author{George V. Eleftheriades}
\affiliation{{The Edward S. Rogers Sr. Department of Electrical and Computer Engineering,
University of Toronto,\\Toronto, ON, Canada M5S 2E4}}
\email{gelefth@ece.utoronto.ca}



\begin{abstract}
\textcolor{black}{Huygens'} metasurfaces are electrically thin devices which allow arbitrary field transformations. Beam refraction is among the first demonstrations of realized metasurfaces. As previously shown for extreme-angle refraction, control over only the electric impedance and magnetic admittance of the \textcolor{black}{Huygens'} metasurface proved insufficient to produce the desired reflectionless field transformation. To maintain zero reflections for wide refraction angles, magnetoelectric coupling between the electric and magnetic response of the metasurface, leading to bianisotropy, can be introduced. In this paper, we report the theory, design, and experimental characterization of a reflectionless bianisotropic metasurface for extreme-angle refraction of a normally incident plane wave towards 71.8 degrees at 20 GHz. The theory and design of three-layer asymmetric bianisotropic unit cells are discussed. The realized printed circuit board (PCB) structure was tested via fullwave simulations as well as experimental characterization. To experimentally verify the prototype, two setups were used. A \textcolor{black}{quasi-optical} experiment was conducted to assess the specular reflections of the metasurface, while a far-field antenna measurement characterized its refraction nature. The measurements verify that the fabricated metasurface has negligible reflections  and the majority of the scattered power is refracted to the desired Floquet mode. This provides an experimental demonstration of a reflectionless wide-angle refracting metasurface using a bianisotropic \textcolor{black}{Huygens'} metasurface at microwave frequencies.

\end{abstract}


\maketitle

\section{Introduction}
\label{sec:introduction}
\textcolor{black}{Huygens'} metasurfaces are 2D equivalents of metamaterials which allow the arbitrary transformation of electromagnetic waves \cite{Pfeiffer2013, Monticone2013, Selvanayagam2013}. The inspiration of \textcolor{black}{Huygens'} metasurfaces originates from the equivalence principle, which demonstrates field transformations via infinitesimal thin surface current densities \cite{Harrington}. By equating the field discontinuities to a set of electric and magnetic currents, the equivalence principle provides the means of engineering boundary conditions for arbitrary wavefront manipulation. \textcolor{black}{Huygens'} metasurfaces can be designed in terms of electric and magnetic impedances or polarizabilities, to generate the required currents from the incident fields to produce the desired wave transformation \cite{Pfeiffer2013,Selvanayagam2013, Achouri2015, Epstein2016_2}. These thin and planar artificial surfaces are composed of subwavelength unit cells, which allow microscopic interactions with incident fields \cite{Holloway2012, Tretyakov2015}. By utilizing these unit cells, metasurfaces can achieve fine spatial sampling of the impedances or polarizabilities to accurately model the theoretical boundary conditions. 

Due to their versatility, metasurface applications have included wave refraction, beam focusing, and polarization control, to name a few \cite{Pfeiffer2013, Monticone2013, Selvanayagam2013, Radi2014, Selvanayagam2014, Kim2014, Wong2015, Pfeiffer2014, Pfeiffer2016_1}. One particularly challenging application of metasurfaces has been wide-angle refraction. While shallow angles of refraction have been demonstrated in the past with good efficiency \cite{Pfeiffer2013, Wong2014}, wide angle refraction with respect to the incident angle, was found to be more challenging. The issue with these surfaces originates from the mismatch of the incident and refracted waves. Due to the different wave impedances seen by the waves on the incident and transmission sides of the refraction metasurface, reflections became more significant as the refraction angle becomes more oblique \cite{Selvanayagam2013, Wong2016}. This issue results from the boundary conditions themselves. As the first \textcolor{black}{Huygens'} refracting metasurfaces only allowed control of the electric impedance and magnetic admittance alone, the resulting realizations were all symmetric \cite{Pfeiffer2013, Monticone2013, Selvanayagam2013, Asadchy2016}. Due to this physical symmetry, it was impossible for these structures to match different input and output wave impedances \cite{Epstein2014_2,Wong2016}. Thus, when wide angle refraction was desired, the surfaces either suffered from large reflections, or required loss and gain regions \cite{Asadchy2016, Asadchy2016_1, Estakhri2016, Estakhri2016_1}. 

The solution to the wide-angle refraction problem was resolved when examining the stipulation in the boundary conditions. As the previous method only allowed control over the electric and magnetic impedances, the  boundary conditions did not allow for a lossless and passive solution. Therefore, an extra degree of freedom was required to resolve this issue. Bianisotropy in the form of a magnetoelectric coupling coefficient was found to be a solution to realizing a lossless and passive surface \cite{Epstein2016_3, Asadchy2016}. The new metasurface formulation accounted for the mismatch problem by modeling the bianisotropy in an asymmetric structure \cite{Epstein2016_3,Chen2017}. The solution was actually first discovered in \cite{Wong2016}, however at the time it was described using a microwave network approach. In \cite{Asadchy2016, Epstein2016_3} it was shown that indeed such an asymmetric structure introduces bianisotropy, providing a coupling mechanism between the electric and magnetic responses of a \textcolor{black}{Huygens'} metasurfaces. This additional degree of freedom can be used to perfectly match the incident and refracted waves even at extreme refraction  angles. Such bianisotropic \textcolor{black}{Huygens'} metasurfaces are ideally lossless, passive and completely reflectionless \cite{Epstein2016_3, Asadchy2016,Chen2017}. 

%

\textcolor{black}{While the theoretical solution of the perfect (reflectionless) refraction was formulated in terms of abstract surface boundary conditions \cite{Epstein2016_3, Asadchy2016}, realization of such devices is far from trivial \textcolor{black}{\cite{Chen2017}}. Implementations of bianisotropic surfaces such as with omega wires \cite{Asadchy2015} and helices \cite{AsadchyPRX2015} have been shown in the past. However, such realizations may not be appealing from an application perspective due to their complexity and fabrication challenges. To alleviate this issue, we proposed to use a 3-layer PCB structure in \textcolor{black}{\cite{Epstein2016_3, Chen2017}} which has shown capabilities of synthesizing \textcolor{black}{Huygens'} metasurfaces in the past \cite{Monticone2013, Epstein2016_2, Pfeiffer2014_3, Epstein2016, PfeifferPRL2014, AchouriArxiv2015}, and is practically simple to realize. However, the viability of these designs still requires experimental validation in the form of a complete design, fabrication, and characterization cycle. Such validation is crucial for the development and realization of more advanced bianisotropic metasurfaces \cite{AbdoSanchezEUCAP2017, EpsteinTAPS2017}. Moreover, in the specific case of wide-angle refraction, the question of whether the bianisotropy that is required to produce reflectionless matching of the incident and refracted waves, can be physically realized still needs to be validated experimentally. While in \cite{LavigneArxiv2017}, experimental characterization of such a perfect refraction surface was conducted, the experimental results which were obtained using near-field techniques could not characterize the reflections of the prototype. Although the results presented therein \cite{LavigneArxiv2017} do demonstrate bianisotropic refraction, a complete experimental verification of perfect (reflectionless) wide-angle refraction, which characterizes both the transmission and the reflections, is yet to be presented.}  

In this paper, we address these issues by presenting a complete design cycle from theory, to design, and to the final experimental verification of a wide-angle reflectionless bianisotropic \textcolor{black}{Huygens'} metasurface. The metasurface refracts a normally incident plane wave towards 71.8$^\circ$ at 20 GHz. \textcolor{black}{The derivation of the boundary conditions will be shown for both non-bianisotropic and bianisotropic cases. While the theoretical boundary conditions have been presented previously in \cite{Epstein2016_3}, they are repeated here for completeness and to demonstrate the benefit of utilizing bianisotropy of such refracting surfaces.} A design strategy which translates the derived boundary conditions to a realizable structure will be detailed. The design of the unit cells will be presented in depth, and fullwave simulation results via Ansys High Frequency Structure Simulator (HFSS) will be shown. \textcolor{black}{The proposed unit cell design is performed without any full-wave optimization of the metasurface period, and the homogenization approach was utilized to reduce the computational effort to individually design the unit cells.} Periodic simulations of one period of the realized metasurface will be conducted and discussed. Finally, a fabricated PCB metasurface will be experimentally verified using a combination of two experimental setups. {A \textcolor{black}{quasi-optical} setup will be described to characterize the reflectionless nature of the proposed metasurface. In addition, a far-field radiation pattern experiment will examine the refractive properties of the surface. The combination of these two measurement techniques allows the characterization of both refracted and reflected fields. With minimal measured reflections for all scattered modes and more than 80\% of the scattered power refracted in the desired direction, this hybrid experimental testing validates the metasurface design. This provides a complete experimental demonstration of a reflectionless wide-angle refracting \textcolor{black}{Huygens'} metasurface.} Moreover, this work verifies the theory and demonstrates the viability of PCB metasurfaces for realizing bianisotropic devices \cite{Asadchy2015,Epstein2016_3,Asadchy2016,Epstein2016_4}.

\section{Theory and Derivation}
\label{sec:theory}

\subsection{Non-bianisotropic Boundary Conditions}
\label{subsec:theory_1}

\begin{figure}[h]
\begin{center}
\vspace{-0.3cm}
\noindent
\includegraphics[scale=0.65]{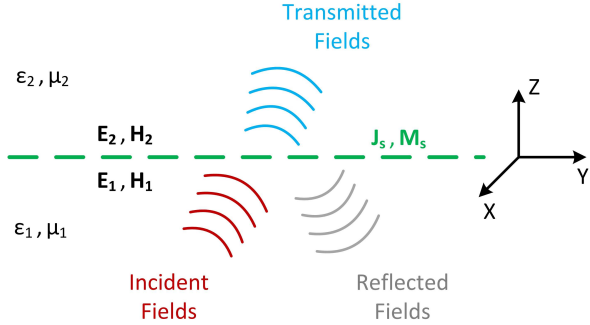}
  \caption{Equivalence principle for arbitrary field transformation.}
  \label{fig:fields0}
 \vspace{-0.6cm}
\end{center}
\end{figure}

\textcolor{black}{Huygens'} metasurfaces utilize the equivalence principle to perform arbitrary wave transformations \cite{Harrington, BalanisEM}. Thus, the first step in designing the \textcolor{black}{Huygens'} metasurface is to derive the boundary conditions required to produce the desired effect. As previously stated, \textcolor{black}{Huygens'} metasurfaces utilize both electric and magnetic currents to model the field discontinuity \cite{Pfeiffer2013, Selvanayagam2013}. This result can readily be seen from the equivalence principle as shown in Eq. ($\ref{eq:bc1}$) and Fig. \ref{fig:fields0}. By stipulating the fields ($\vec{{E}}_{1}$, $\vec{{H}}_{1}$ and $\vec{{E}}_{2}$, $\vec{{H}}_{2}$) in two half spaces, the necessary tangential electric ($\vec{{J}}_{{s}}$) and magnetic ($\vec{{M}}_{{s}}$) currents required for the field transformation can be obtained. Therefore, if the necessary current could be excited, the field transformation would occur as desired. 

However, the equivalence principle does not state how the currents can be physically realized. One method is to use impressed sources in space to generate these currents \cite{Selvanayagam2012,Selvanayagam2013_1}. \textcolor{black}{In this approach, active current sources, both electric and magnetic, must be used.} However, introducing active sources which must generate spatially varying current densities is not a trivial task. Another approach is to utilize the fields themselves to excite the currents. Utilizing this approach, a secondary relation between the fields and currents can be introduced. In this method, instead of generating the currents directly, properties of the metasurface can be designed in order to induce the desired currents from the incident fields \cite{Kuester2003}. Similar to how an electric current density can be related to an applied electric field by the electric conductivity, we can introduce an electric impedance which can relate the average tangential electric field on the metasurface to the surface electric current. Similarly, a magnetic admittance can relate the average magnetic fields to the magnetic current density \cite{Selvanayagam2013,Pfeiffer2013,Kuester2003}. 

In general the electric impedance and magnetic admittance are tensors accounting for arbitrary polarization of the desired fields \cite{Pfeiffer2016_1,Selvanayagam2014, Epstein2016_2}. For our demonstration, the 1D refraction of transverse electric (TE) waves, meaning that the electrical field component will always be tangential to the boundary of the field discontinuity (or perpendicular to the plane of incidence), will be presented. Due to the choice of polarization, the electric impedance and magnetic admittance can then be represented as scalar quantities \cite{Epstein2016_3, Epstein2016_2}. The generalized TE field quantities can be seen in Eq. (\ref{eq:bc_A1}), where $\vec{{E}}_{1}$, $\vec{{H}}_{1}$ and $\vec{{E}}_{2}$, $\vec{{H}}_{2}$ denote the total fields in the two regions (see axes in Fig. \ref{fig:fields0}). Applying the field profiles to the equivalence principle in Eq. (\ref{eq:bc1}) at a desired boundary, the required electric impedance and magnetic admittance can be found. In this case, the boundary will be the z=0 plane with $\vec{{E}}_{1}^{-}$, $\vec{{H}}_{1}^{-}$ and $\vec{{E}}_{2}^{+}$, $\vec{{H}}_{2}^{+}$ representing the fields at the surface of the boundary as shown in Eq. (\ref{eq:bc_A2}). The electric impedance ${Z_{se}}$ and magnetic admittance ${Y_{sm}}$ can then be related to the fields as seen in Eq. ($\ref{eq:bc2}$), where $\vec{{E}}_{{t,\mathrm{avg}}}$ and $\vec{{H}}_{{t,\mathrm{avg}}}$ are the average tangential fields at the boundary as shown in Eq. (\ref{eq:bc_A3}) \cite{Kuester2003,Epstein2016_3}. Combining Eq. ($\ref{eq:bc1}$) and Eq. ($\ref{eq:bc2}$) together and applying the boundary field components, we can then form a system of equations as shown in Eq. ($\ref{eq:bc3}$) and Eq. ($\ref{eq:bc4}$) \cite{Epstein2014}, which relates the desired field quantities to the properties of the metasurface in the form of the electric impedance and the magnetic admittance \cite{Pfeiffer2013, Selvanayagam2013}. By simple manipulation of these two equations, the surface impedance/admittance boundary conditions can then be solved uniquely as shown in Eq. ($\ref{eq:bc5}$) \cite{Epstein2014}.

\begin{equation} \label{eq:bc1}
\vec{J}_{s} = \hat{n} \times (\vec{H}^+_2-\vec{H}_1^-) \hspace{1cm} \vec{M}_{s} = -\hat{n} \times (\vec{E}_2^+-\vec{E}_1^-)
\end{equation}

\begin{equation} \label{eq:bc_A1}
\begin{cases}
\vec{E}_{1} = E_{1,x} \hat{x} \hspace{0.15cm},\hspace{0.15cm} \vec{H}_{1} = H_{1,y} \hat{y} + H_{1,z} \hat{z}\\
\vec{E}_{2} = E_{2,x} \hat{x} \hspace{0.15cm},\hspace{0.15cm} \vec{H}_{2} = H_{2,y} \hat{y} + H_{2,z} \hat{z}
\end{cases}
\end{equation}

\begin{equation} \label{eq:bc_A2}
\begin{cases}
\vec{E}_{1}^{-} = \vec{E}_{1}(y,z \rightarrow 0^-) \hspace{0.15cm},\hspace{0.15cm} \vec{H}_{1}^{-} = \vec{H}_{1}(y,z\rightarrow0^-)\\
\vec{E}_{2}^{+} = \vec{E}_{2}(y,z\rightarrow0^+) \hspace{0.15cm},\hspace{0.15cm} \vec{H}_{2}^{+} = \vec{H}_{2}(y,z\rightarrow0^+)
\end{cases}
\end{equation}

\begin{equation} \label{eq:bc2}
\vec{E}_{t,\mathrm{avg}} = Z_{{se}}\vec{J}_{s} \hspace{1cm} \vec{H}_{t,\mathrm{avg}} = Y_{{sm}}\vec{M}_\mathrm{s}
\end{equation}

\begin{equation} \label{eq:bc_A3}
\vec{E}_{t,\mathrm{avg}} = \frac{1}{2}(\vec{E}_{{1,t}}^{-} + \vec{E}_{{2,t}}^{+}) \hspace{0.7cm}
\vec{H}_{t,\mathrm{avg}} = \frac{1}{2}(\vec{H}_{{1,t}}^{-} + \vec{H}_{{2,t}}^{+})
\end{equation}

\begin{equation} \label{eq:bc3}
\frac{1}{2}(\vec{E}_{{1,t}}^{-}+\vec{E}_{{2,t}}^{+}) = Z_{{se}}[\hat{n} \times (\vec{H}_2^{+}-\vec{H}_{1}^{-})] 
\end{equation}

\begin{equation} \label{eq:bc4}
\frac{1}{2}(\vec{H}_{{1,t}}^{-}+\vec{H}_{{2,t}}^{+}) = Y_{{sm}}[- \hat{n} \times (\vec{E}_2^{+}-\vec{E}_1^{-})]
\end{equation}




\begin{equation} \label{eq:bc5}
Z_{{se}} = \dfrac{E_{1,x}^{-}+E_{2,x}^{+}}{2(H_{1,y}^{-}-H_{2,y}^{+})} \hspace{0.25cm},\hspace{0.25cm} Y_{{sm}} = \dfrac{H_{1,y}^{-}+H_{2,y}^{+}}{2(E_{1,x}^{-}-E_{2,x}^{+})}
\end{equation}

\begin{figure}[h]
\begin{center}
\vspace{-0.3cm}
\noindent
\includegraphics[scale=0.65]{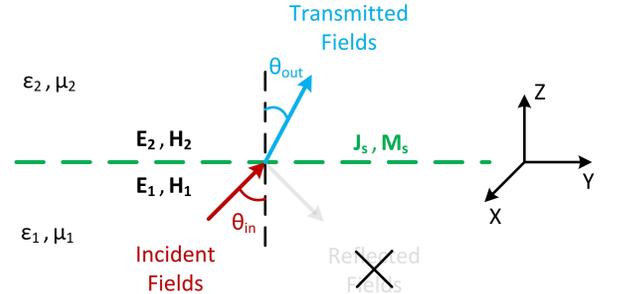}
  \caption{Desired wave components for reflectionless refraction.}
  \label{fig:fields1}
 \vspace{-0.6cm}
\end{center}
\end{figure}

Applying the above derivations, the surface boundary conditions required for any arbitrary TE field transformation can be obtained. Therefore, by stipulating the desired fields and applying the boundary conditions properly, the corresponding metasurface can be designed. In the case of reflectionless refraction, the desired waves are illustrated in Fig. $\ref{fig:fields1}$. In traditional materials, the refraction angle will follow Snell's law and the reflections are determined by the material properties and the angle of incidence \cite{BalanisEM}. However, in the case of the desired \textcolor{black}{metasurface}, the refraction angle can be arbitrarily set and the reflections can be stipulated to vanish. Therefore, by setting the reflected fields to be identically zero, the surface boundaries should produce the necessary impedances/admittances to produce a reflectionless structure. On the incident domain then, the total fields can be described by $\vec{{E}}_\mathrm{1}$ and $\vec{{H}}_\mathrm{1}$, which corresponds to the incident wave. While on the transmission side, the refracted fields are described as $\vec{{E}}_\mathrm{2}$ and $\vec{{H}}_\mathrm{2}$. Referring to Fig. $\ref{fig:fields1}$, the incident wave will then impinge on the surface with angle ${\theta_\mathrm{in}}$ and the transmitted fields will depart from the surface with angle ${\theta_\mathrm{out}}$. Using these field stipulations, we can then apply them to a specific refraction scenario.

\begin{figure*}
\begin{center}
\vspace{-0.3cm}
\noindent
\includegraphics[scale=0.48]{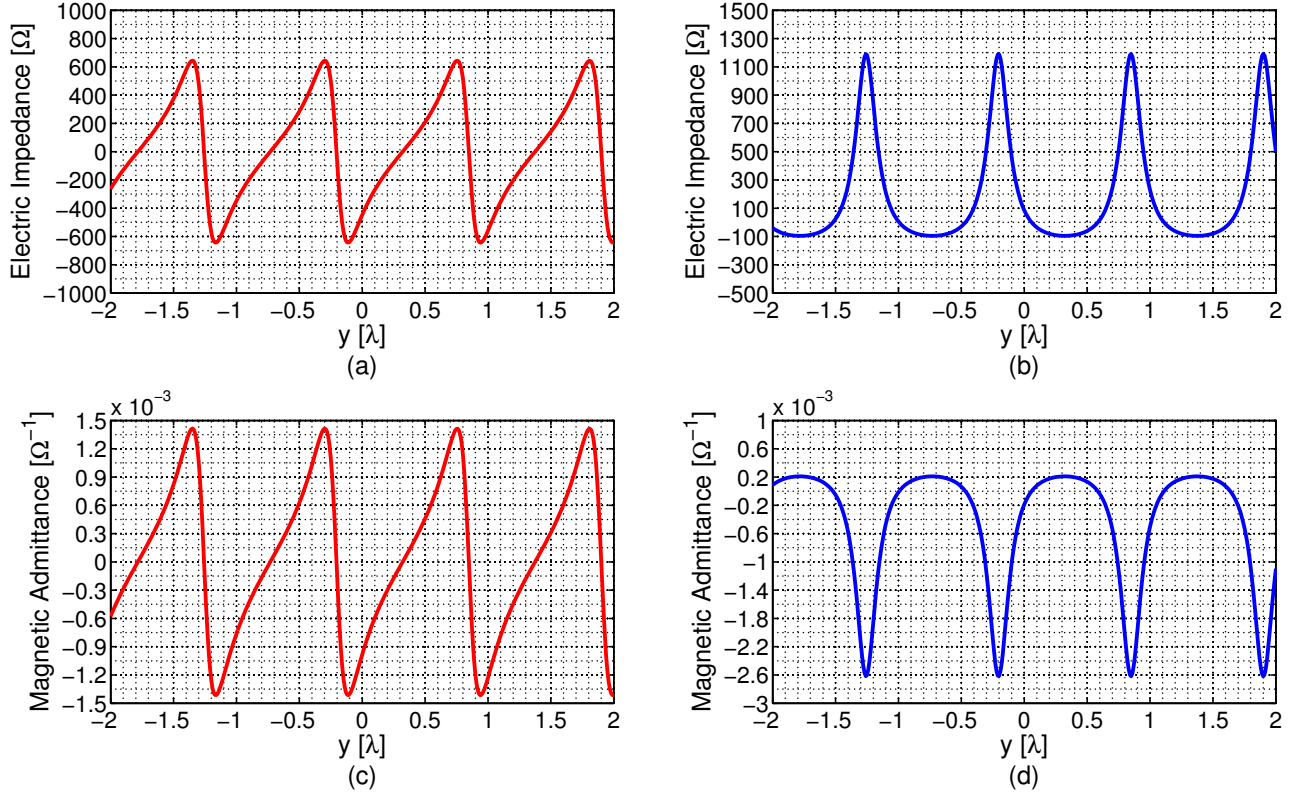}
  \caption{\textcolor{black}{Non-bianisotropic boundary conditions for 0$^\circ$ to 71.8$^\circ$ refraction at 20 GHz. a) Imaginary component of Z$_{{se}}$. b) Real component of Z$_{{se}}$. c) Imaginary component of Y$_{{sm}}$. d) Real component of Y$_{{sm}}$.}}
  \label{fig:monoimpede1}
 \vspace{-0.6cm}
\end{center}
\end{figure*}

For the demonstration of reflectionless refraction, the chosen refraction scenario will be at 20 GHz with a normally incident plane wave refracting towards 71.8$^\circ$. As previous mentioned, both the incident and transmitted fields will be TE polarized, meaning that the electrical field component will always be tangential to the boundary of the field discontinuity (or perpendicular to the plane of incidence). In the case of this TE refraction, the field profiles can be written in the general form as shown in Eq. ($\ref{eq:fields1}$) and Eq. ($\ref{eq:fields2}$), representing the  components of the incident and refracted waves that are tangential to the interface. Here $k_0$ and $\eta$ denote the wave number and wave impedance of free space, and $Z_{0,1}$ and $Z_{0,2}$ denote the wave impedance of the incident and transmitted fields respectively. By choosing the incident angles ${\theta_\mathrm{in}}$ and ${\theta_\mathrm{out}}$, arbitrary refraction can be achieved. In this case, ${\theta_\mathrm{in}}$ will be 0$^\circ$ corresponding to the normally incident plane wave and ${\theta_\mathrm{out}}$ will be 71.8$^\circ$ corresponding to the refracted output wave. Additionally, to derive the necessary transmission wave magnitude, local power conservation conditions must be applied as shown in Eq. ($\ref{eq:power_conserv1}$) \cite{Epstein2016_3}. By imposing this condition, we stipulate that we desire to transfer all the power from the incident wave to the refracted beam. Through applying this condition, the refracted wave magnitude can be determined as seen in Eq. ($\ref{eq:power_conserv2}$). Using these field distributions and Eq. ($\ref{eq:bc5}$) the refraction boundary conditions are calculated and shown in Fig. $\ref{fig:monoimpede1}$.

\begin{equation} \label{eq:fields1}
\begin{cases}
E_{x,1}(y,z) = E_{0,1}e^{-jk_0{\cos}\theta_{\mathrm{in}}z}e^{-jk_0{\sin}\theta_{\mathrm{in}}y} \\
H_{y,1}(y,z) = \dfrac{1}{Z_{0,1}}E_{0,1}e^{-jk_0{\cos}\theta_{\mathrm{in}}z}e^{-jk_0{\sin}\theta_{\mathrm{in}}y} \\
Z_{0,1} = \dfrac{\eta}{{\cos}\theta_{\mathrm{in}}}
\end{cases}
\end{equation}

\begin{equation} \label{eq:fields2}
\begin{cases}
E_{x,2}(y,z) = E_{0,2}e^{-jk_0{\cos}\theta_{\mathrm{out}}z}e^{-jk_0{\sin}\theta_{\mathrm{out}}y} \\
H_{y,2}(y,z) = \dfrac{1}{Z_{0,2}}E_{0,2}e^{-jk_0{\cos}\theta_{\mathrm{out}}z}e^{-jk_0{\sin}\theta_{\mathrm{out}}y} \\
Z_{0,2} = \dfrac{\eta}{{\cos}\theta_{\mathrm{out}}}
\end{cases}
\end{equation}

\begin{equation} \label{eq:power_conserv1}
P_{z,1} = \dfrac{1}{2}\Re \lbrace E_{x,1}^{-}H_{y,1}^{-*} \rbrace = \dfrac{1}{2}\Re \lbrace E_{x,2}^{+}H_{y,2}^{+*} \rbrace = P_{z,2}
\end{equation}

\begin{equation} \label{eq:power_conserv2}
|E_{0,2}| = \sqrt{\dfrac{Z_{0,2}}{Z_{0,1}}} |E_{0,1}|
\end{equation}

As shown, the calculated electric impedance and magnetic admittance are both complex and contain non-zero real and imaginary components \cite{Asadchy2016}. However, this proves to be problematic as the non-zero real components of these impedances and admittances lead to boundary conditions which require controlled loss and gain mechanisms. As previously mentioned, this is actually a well known outcome \cite{Asadchy2016, Asadchy2016_1, Estakhri2016}. However, investigating the boundary conditions, the fundamental issue can be uncovered and correspondingly alleviated. Examining Eq. ($\ref{eq:bc3}$) and Eq. ($\ref{eq:bc4}$), it is evident that the complex system of equations has two sets of complex unknowns, ${Y_{sm}}$ and ${Z_{se}}$ and two sets of complex equations. Mathematically, this will lead to a unique solution for any given field transformation. Thus, once the desired input and output fields are stipulated, the solution is uniquely determined without any constraint on the passivity and/or losslessness of the field transformation. Therefore, to circumvent this problem, additional degrees of freedom must be introduced. A solution is then to introduce another unknown into the boundary conditions.

Investigating Eq. ($\ref{eq:bc1}$) and Eq. ($\ref{eq:bc2}$), it is clear that Eq. ($\ref{eq:bc1}$) cannot be altered as it represents the equivalence principle. However, Eq. ($\ref{eq:bc2}$) can certainly be modified, which would restructure how the surface currents can be related to the fields. In the current method, the electric field will excite an electric current and the magnetic field will excite a magnetic current. However, no cross excitation exists. Therefore, to introduce more unknowns to the problem, a simple solution is to add bianisotropy which amounts to the coupling of the electric and magnetic currents in the boundary conditions. In this fashion, bianisotropy allows both the tangential electric and magnetic fields to independently excite both electric and magnetic currents. By introducing a bianisotropic magnetoelectric coupling coefficient, an additional degree of freedom is introduced, thus allowing the synthesis of a lossless, passive, and reflectionless metasurface \cite{Epstein2016_3, Asadchy2016,Chen2017}.

\subsection{Bianisotropic Boundary Conditions}
\label{subsec:theory_2}


To involve bianisotropy into the formulation, Eq. ($\ref{eq:bc2}$) can be modified by introducing a magnetoelectric coupling coefficient ${K_{em}}$ as seen in Eq. ($\ref{eq:bibc1}$) and Eq. ($\ref{eq:bibc10}$) \cite{Epstein2016_3}. This coupling coefficient physically represents a coupling of the the electric and magnetic fields \cite{Kong1972}. The outcome is that both electric and magnetic fields may excite both electric and magnetic currents \cite{Epstein2016_3}. By combining Eq. ($\ref{eq:bibc1}$), Eq. ($\ref{eq:bibc10}$), and the unaltered equivalence principle in Eq. ($\ref{eq:bc1}$), the new TE bianisotropic boundary conditions can be formulated as seen in Eq. ($\ref{eq:bibc2}$) and Eq. ($\ref{eq:bibc3}$) \cite{Epstein2016_3}. 

\begin{equation} \label{eq:bibc1}
\vec{E}_{t,\mathrm{avg}} = Z_{{se}}\vec{J}_{s}   - K_{{em}}[\hat{n}\times \vec{M}_{s}]
\end{equation}

\begin{equation} \label{eq:bibc10}
\vec{H}_{t,\mathrm{avg}} = Y_{{sm}}\vec{M}_{s}  - K_{{em}}[\hat{n}\times \vec{J}_{s}]
\end{equation}


\begin{eqnarray}\label{eq:bibc2}
\frac{1}{2}(\vec{E}_{{1,t}}^{-}+\vec{E}_{{2,t}}^{+}) &&= Z_{{se}}[\hat{n} \times (\vec{H}_2^{+}-\vec{H}_1^{-})] \\ \nonumber
&&- K_{{em}}\{\hat{n}\times [-\hat{n} \times (\vec{E}_2^{+}-\vec{E}_1^{-})]\} 
\end{eqnarray} 

\begin{eqnarray}\label{eq:bibc3}
\frac{1}{2}(\vec{H}_{{1,t}}^{-}+\vec{H}_{{2,t}}^{+}) &&= Y_{{sm}}[- \hat{n} \times (\vec{E}_2^{+}-\vec{E}_1^{-})] \\ \nonumber
 &&- K_{{em}}\{\hat{n}\times [\hat{n} \times (\vec{H}_2^{+}-\vec{H}_1^{-})] \} 
\end{eqnarray}

Examining the new set of bianisotropic boundary conditions, it is clear that by involving the coupling coefficient ${K_{em}}$, a new degree of freedom has indeed been introduced. While the boundary conditions are still encased in a system of two complex equations, there now exist three complex unknowns being the electric impedance ${Z_{se}}$, the magnetic admittance ${Y_{sm}}$, and the magnetoelectric coupling coefficient ${K_{em}}$. Due to the increased number of unknowns, the problem is essentially over specified. However, this redundancy now allows the specification of a lossless and passive solution. 

\begin{figure*}
\begin{center}
\vspace{-0.3cm}
\noindent
\includegraphics[scale=0.48]{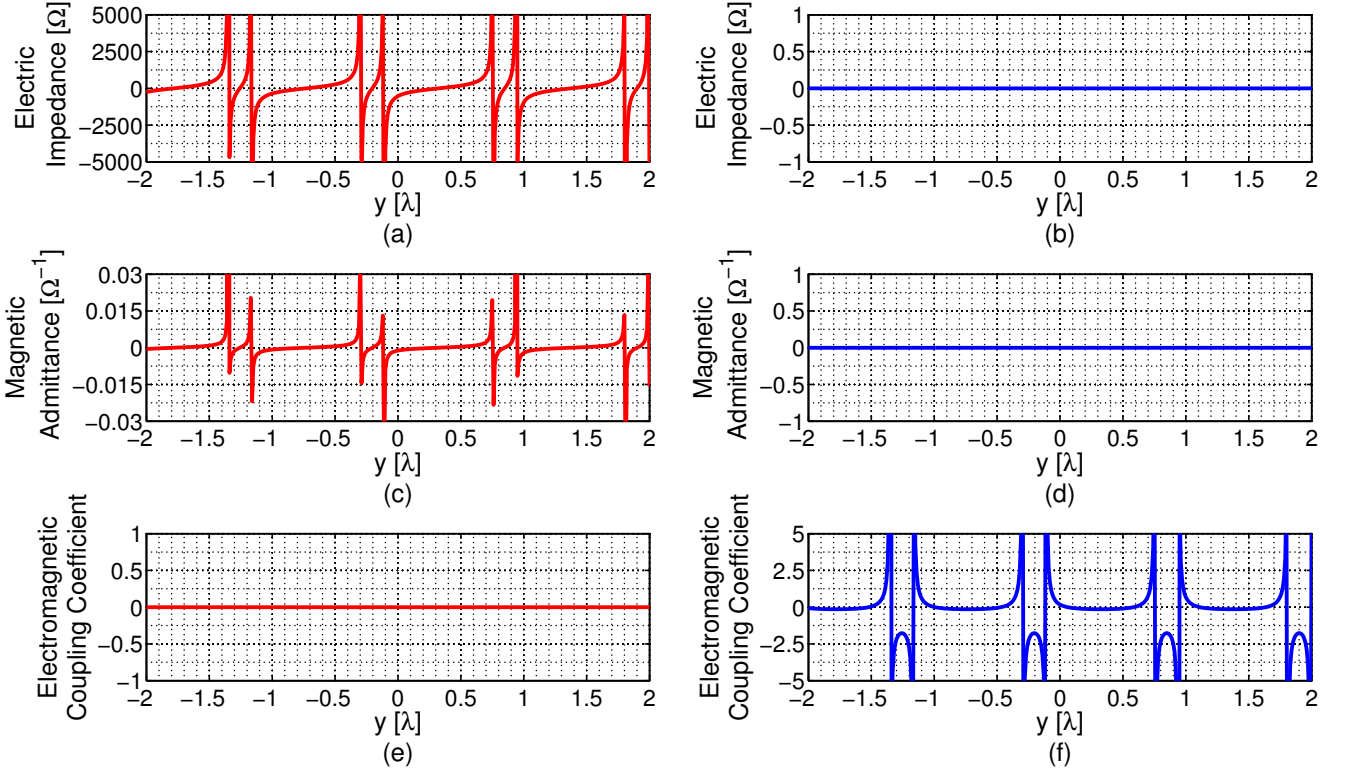}
  \caption{\textcolor{black}{Bianisotropic boundary conditions for 0$^\circ$ to 71.8$^\circ$ refraction at 20 GHz. a) Imaginary component of Z$_{{se}}$. b) Real component of Z$_{{se}}$. c) Imaginary component of Y$_{{sm}}$. d) Real component of Y$_{{sm}}$. e) Imaginary component of K$_{{em}}$. f) Real component of K$_{{em}}$.}}
  \label{fig:biimpede}
 \vspace{-0.6cm}
\end{center}
\end{figure*}

As seen previously in Fig. $\ref{fig:monoimpede1}$, the real component of the electric impedance and magnetic admittance are non-zero for the non-bianisotropic boundary conditions. The obvious choice for the bianisotropic solution would then be to set the impedance conditions to have vanishing real components. Once these two values are specified, the redundancy in the solution is reduced, and the resulting components of the unknowns can again be uniquely determined. In general, the solution for the new bianisotropic conditions will result in ${K_{em}}$ having both real and imaginary components. However, once the local power conservation condition in Eq. (\ref{eq:power_conserv1}) is applied to the field distributions similar to Sec. \ref{subsec:theory_1}, the resulting ${K_{em}}$ value will only contain real components \cite{Epstein2016_3}. Therefore, while in general ${K_{em}}$ is a complex parameter, in our specific case its imaginary component will vanish, which corresponds to a passive and lossless implementation \cite{Radi2013,Epstein2016_3}. For simplicity, the resulting analytical solutions to the lossless and passive bianisotropic boundary conditions specific to our local power conservation condition can be seen in Eq. ($\ref{eq:bibc5}$), Eq. ($\ref{eq:bibc6}$), and Eq. ($\ref{eq:bibc7}$) \cite{Epstein2016_3}. The refraction problem can then be applied and the correspondingly ${Z_{se}}$, ${Y_{sm}}$, and ${K_{em}}$ values can be seen in Fig. $\ref{fig:biimpede}$.

\begin{eqnarray}\label{eq:bibc5}
Z_{{se}} = -j&&\left[\dfrac{1}{2}  \Im{\left\lbrace \dfrac{E_{1,x}^{-}+E_{2,x}^{+}}{H_{2,y}^{+}-H_{1,y}^{-}} \right\rbrace}\right] \\ \nonumber
&& -j\left[K_{{em}}\Im{\left\lbrace \dfrac{E_{2,x}^{+}-E_{1,x}^{-}}{H_{2,y}^{+}-H_{1,y}^{-}} \right\rbrace}\right] 
\end{eqnarray} 

\textcolor{black}{\begin{eqnarray}\label{eq:bibc6}
Y_{{sm}} = -j&&\left[\dfrac{1}{2}\Im{\left\lbrace \dfrac{H_{1,y}^{-}+H_{2,y}^{+}}{E_{2,x}^{+}-E_{1,x}^{-}} \right\rbrace}\right]  \\ \nonumber
&&+j\left[K_{{em}}\Im{\left\lbrace \dfrac{H_{2,y}^{+}-H_{1,y}^{-}}{E_{2,x}^{+}-E_{1,x}^{-}} \right\rbrace}\right]
\end{eqnarray} }

\begin{equation} \label{eq:bibc7}
K_{{em}} = \dfrac{1}{2}\dfrac{\Re\lbrace{E_{2,x}^{+}H_{1,y}^{-*} - E_{1,x}^{-}H_{2,y}^{+*} \rbrace}}{\Re\lbrace{(E_{2,x}^{+}-E_{1,x}^{-})(H_{2,y}^{+}-H_{1,y}^{-})^*\rbrace}}
\end{equation}

%
%

As seen in Fig. $\ref{fig:biimpede}$, due to the stipulation for vanished real components, the electric impedance and magnetic admittance values are now purely imaginary. \textcolor{black}{Additionally, as previously mentioned, accounting for the desired local power conservation condition, the magnetoelectric coupling coefficient is purely real.} Combining these effects, the new bianisotropic boundary conditions no longer require gain and loss mechanisms, and thus a true lossless and passive solution has been achieved. With the bianisotropic boundary conditions now obtained, realization of the metasurface can be carried out.

\section{Design and Physical Realization}
\label{sec:design}

\subsection{Impedance Matrix Translation}
\label{subsec:design_1}

Using the previously determined boundary conditions for the desired refraction scenario, the metasurface can now be physically realized. However, while the boundary conditions relate the surface currents to the desired fields, it is not intuitive how the physical metasurface can be designed. To assist with the physical realization, the field boundary conditions can be translated into an impedance equivalent system. 

\begin{figure}[h]
\begin{center}
\vspace{-0.3cm}
\noindent
\includegraphics[scale=0.55]{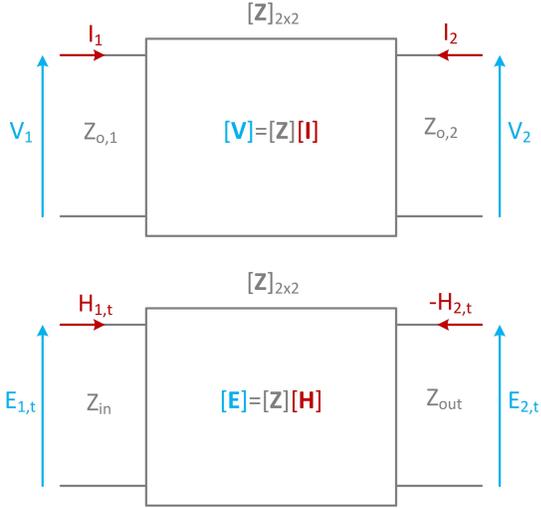}
  \caption{Microwave network equivalence transformation of field boundary conditions.}
  \label{fig:zmatrix}
 \vspace{-0.6cm}
\end{center}
\end{figure}

Taking inspiration from microwave network theory, it is well known that any 2-port microwave device can be characterized in terms of an impedance matrix \cite{Pozar}. Specifically, the impedance or Z matrix relates the currents and voltages applied to the ports of the device. In the case of translating field boundary conditions to an equivalent Z matrix, the ports of the metasurface can be thought of as the two half spaces containing the desired fields. One port of the equivalent network will represent the domain of the normally incident plane wave while the second port will resemble the domain of the refracted plane wave \cite{Wong2016}. In this fashion, the excitation of the ports will be the corresponding electric and magnetic fields as opposed to voltages and currents as in standard microwave circuits. As the electric field is analogous to a voltage and the magnetic field is analogous to a current, the boundary conditions can be rearranged into a matrix format to resemble that of a microwave 2-port network as illustrated in Fig. $\ref{fig:zmatrix}$ \cite{Wong2016, Epstein2016_3}. By shuffling the boundary conditions as seen in Eq. ($\ref{eq:matrix1}$) and Eq. ($\ref{eq:matrix2}$), a matrix format of the boundary conditions can be cast in Eq. ($\ref{eq:matrix3}$) \cite{Epstein2016_3,Selvanayagam2014}.

%

\begin{eqnarray}\label{eq:matrix1}
(\frac{1}{2}-K_{{em}})E_{1,x}^{-} +&& (\frac{1}{2}+K_{{em}})E_{2,x}^{+}\\ \nonumber
 &&= Z_{{se}}H_{1,y}^{-}-Z_{{se}}H_{2,y}^{+}
\end{eqnarray}

\begin{eqnarray} \label{eq:matrix2}
-Y_{{sm}}E_{1,x}^{-}+Y_{{sm}}E_{2,x}^{+} &&= (-K_{{em}}-\frac{1}{2})H_{1,y}^{-} \\ \nonumber
&&+(K_{{em}}-\frac{1}{2})H_{2,y}^{+}
\end{eqnarray}

\textcolor{black}{
\begin{equation} \label{eq:matrix3}
\left[ \begin{tabular}{c}
$E_{1,x}^{-}$\\
$E_{2,x}^{+}$
\end{tabular}\right] 
= \left[ \begin{tabular}{cc}
$Z_{11}$ & $Z_{12}$\\
$Z_{21}$ & $Z_{22}$
\end{tabular}\right]
\left[ \begin{tabular}{c}
$H_{1,y}^{-}$\\
$-H_{2,y}^{+}$
\end{tabular}\right]
\end{equation}}

\begin{equation} \label{eq:matrix4}
Z_{11} = Z_{{se}}+\dfrac{(1+2K_{{em}})^2}{4Y_{{sm}}} \hspace{0.1cm},\hspace{0.1cm} Z_{22} = Z_{{se}}+\dfrac{(1-2K_{{em}})^2}{4Y_{{sm}}}
\end{equation}

\begin{equation} \label{eq:matrix5}
Z_{12} = Z_{21} = Z_{{se}}-\dfrac{(1+2K_{{em}})(1-2K_{{em}})}{4Y_{{sm}}}
\end{equation}

\begin{figure*}
\begin{center}
\vspace{-0.3cm}
\noindent
\includegraphics[scale=0.48]{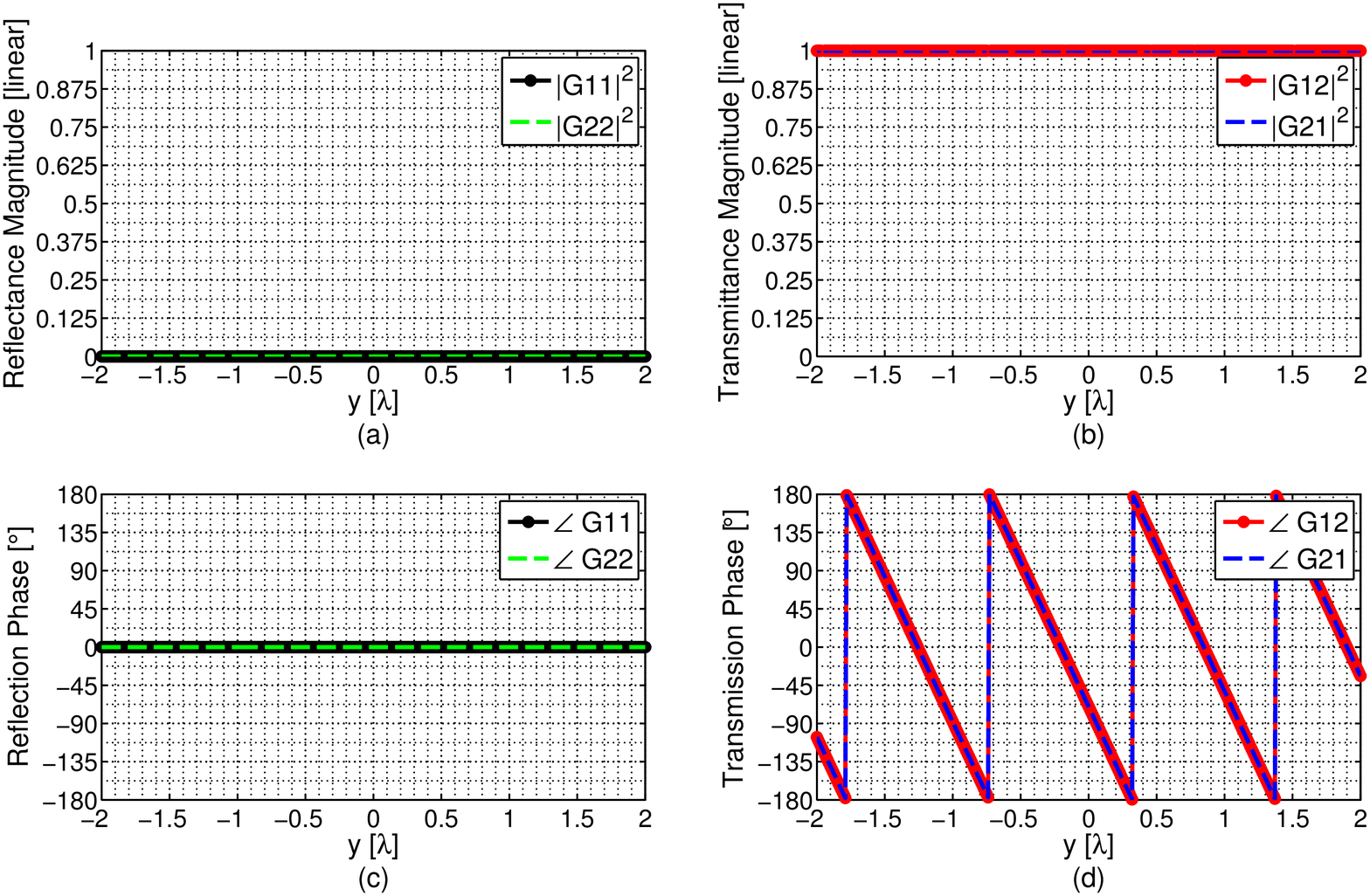}
  \caption{G parameters for bianisotropic refraction of 0$^\circ$ to 71.8$^\circ$ at 20 GHz. a) Magnitude of G$_{11}$ and G$_{22}$. b) Magnitude of G$_{12}$ and G$_{21}$. c) Phase of G$_{11}$ and G$_{22}$. d) Phase of G$_{12}$ and G$_{21}$}
  \label{fig:biS}
 \vspace{-0.6cm}
\end{center}
\end{figure*}

Once in this matrix format, the boundary conditions can then be related to the Z parameters in an equivalent Z matrix as seen in Eq. ($\ref{eq:matrix4}$) and Eq. ($\ref{eq:matrix5}$) \cite{Epstein2016_3}. By applying this Z matrix translation, the field boundary conditions can now be examined as a microwave 2-port network. Using this technique, the translated Z matrix allows us to view the boundary conditions in many perspectives. \textcolor{black}{One useful perspective is to translate the equivalent matrix into its corresponding generalized scattering or G matrix \cite{Frickey1994, Wong2016}. As the metasurface is designed for refraction, the different angles of incidence of the stipulated fields translate to different port impedances in the microwave equivalence. Thus, the generalized scattering parameters are used. Comparatively, the standard S parameters are more difficult to utilize as the S matrix loads both the incident and refraction ports with the same port impedance, which do not represent well the physical phenomenon. In this case, as the desired refraction is for a normally incident TE plane wave refracting to 71.8$^\circ$, the Z matrix can be transformed into the desired G parameters by applying the transformation detailed in \cite{Frickey1994}, and setting the normally incident port impedance to that of free space, $\sqrt{\frac{\mu_0}{\epsilon_0}}=377 \Omega$ and the refraction port impedance to $\frac{377}{\cos(71.8^\circ)} \approx 1207 \Omega$, which is the wave impedance of a TE-polarized plane wave propagating towards 71.8$^\circ$ \cite{Wong2016, Epstein2016_3}. Using these port impedance values, the desired G matrix values for our refraction can be obtained and are presented in Fig. $\ref{fig:biS}$.}

Examining the G matrix parameters, it is clear that G$_{11}$ and G$_{22}$, which correspond to the port reflections, are zero. This proves that indeed our boundary conditions are able to produce reflectionless wave transformations. Additionally, the magnitudes of G$_{21}$ and G$_{12}$, which implement the field transformation, are both unity. This shows that the refraction is performed with no losses. Furthermore, the G$_{21}$ phase is linear and corresponds to the phase required for the refraction. By examining this microwave equivalence system, a more intuitive understanding of the boundary conditions is obtained. Moreover, this translation technique is also useful in the metasurface physical realization. 

\begin{figure}[h]
\begin{center}
\vspace{-0.3cm}
\noindent
\includegraphics[scale=0.35]{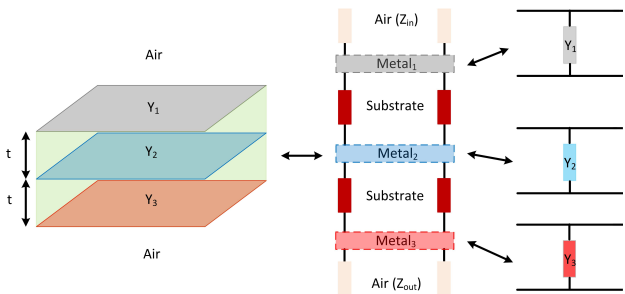}
  \caption{Proposed unit cell comprising 3  shunt admittances separated by dielectric layers.}
  \label{fig:layer}
 \vspace{-0.6cm}
\end{center}
\end{figure}

\begin{figure}
\captionsetup[subfigure]{labelformat=empty}
\begin{center}
\vspace{-0.3cm}
\noindent
\subfloat[]{\includegraphics[scale=0.22]{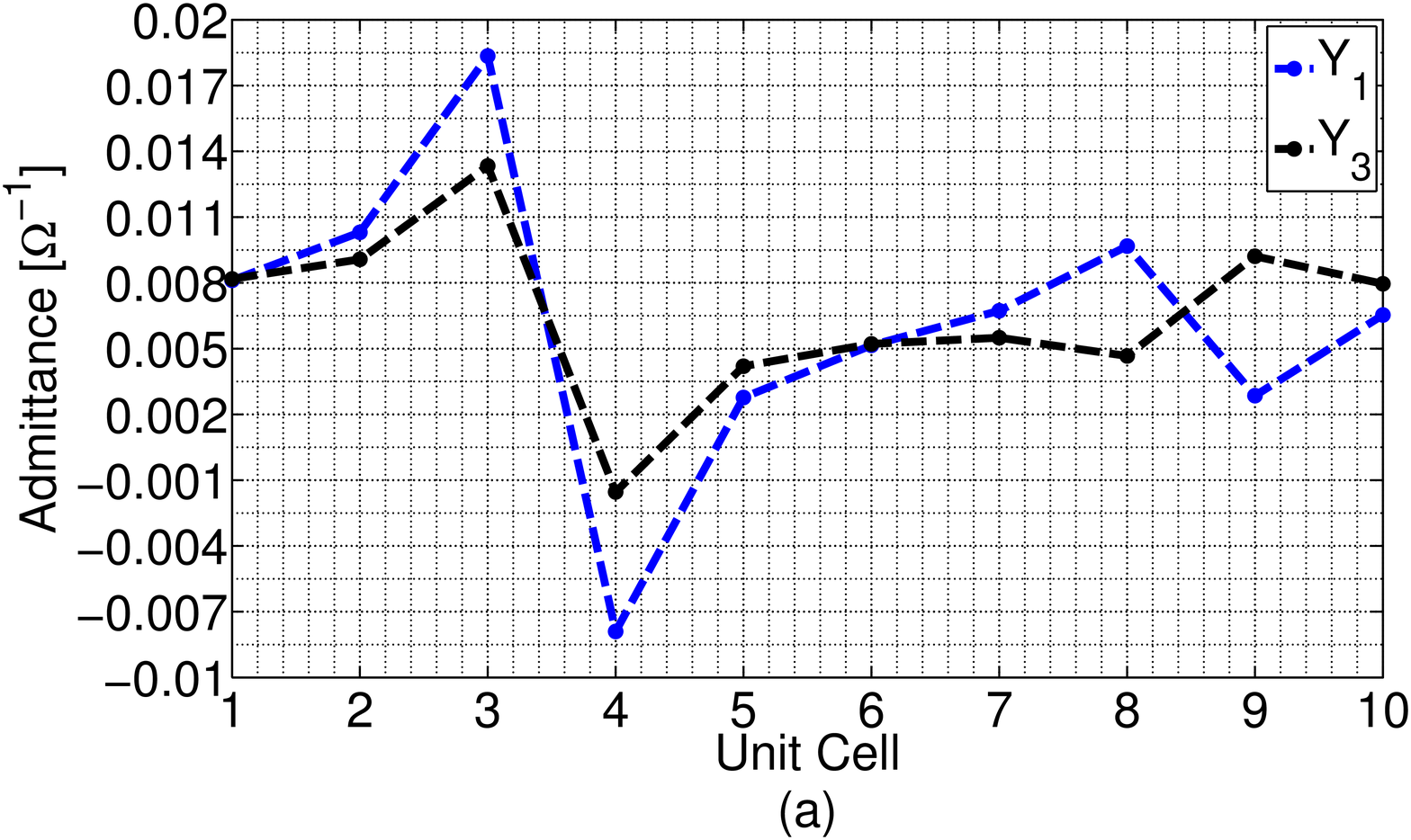}}

\subfloat[]{\includegraphics[scale=0.22]{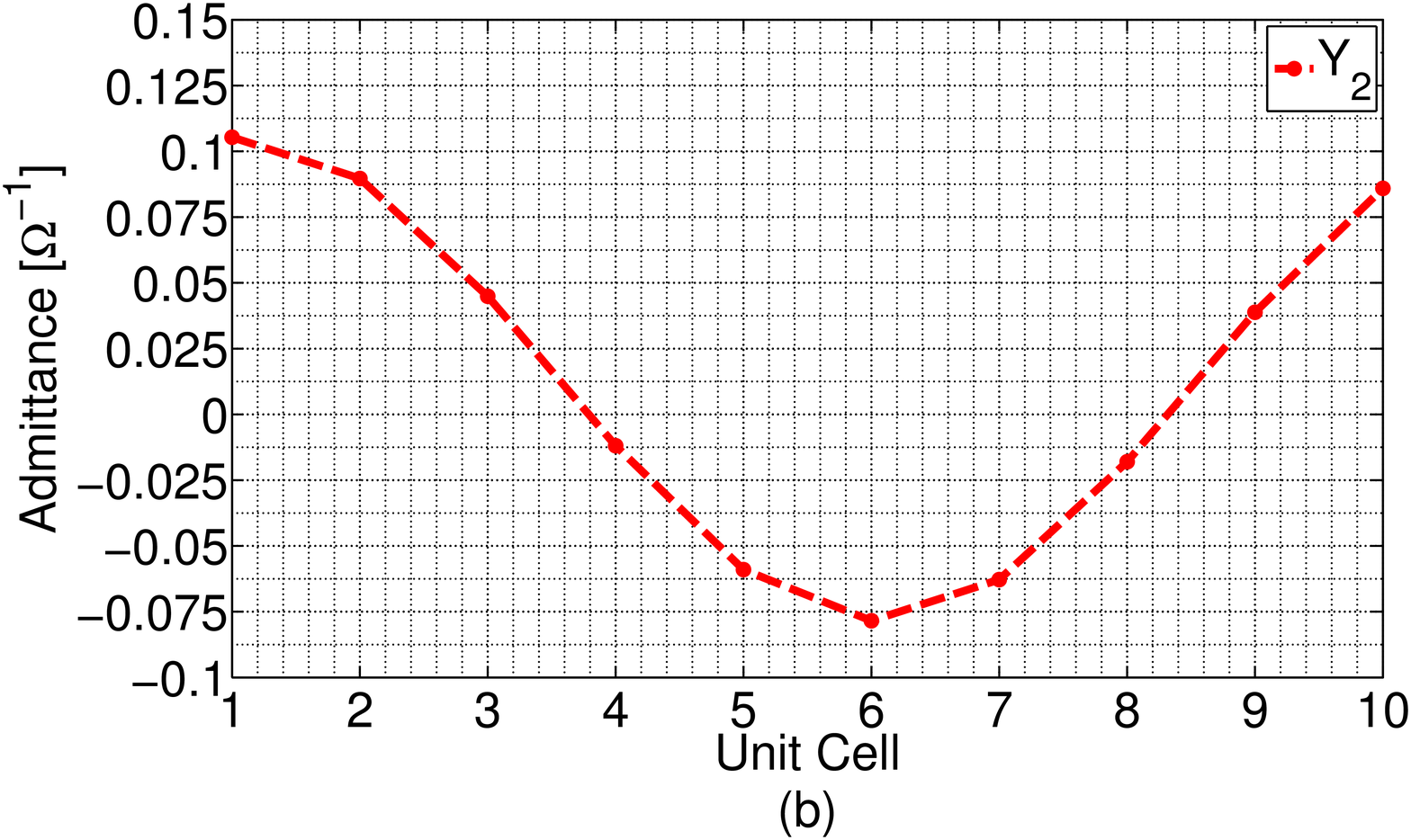}}
  \caption{Proposed unit cell shunt admittances for bianisotropic refraction of 0$^\circ$ to 71.8$^\circ$ at 20 GHz. a) Shunt admittance of outer layers Y$_1$ and Y$_3$. b) Shunt admittance of middle layer Y$_2$.}
  \label{fig:theoryimpede}
 \vspace{-0.6cm}
\end{center}
\end{figure}

To demonstrate the effectiveness of using the equivalent Z matrix in realizing the metasurface, the boundary conditions are first discretized to be spatially sampled. Each discretization point will represent a unit cell that needs to model the boundary condition at that spatial location. \textcolor{black}{Through idealized HFSS simulations using impedance/admittance sheets, we have determined that 10 unit cells per period yield good transmission with low reflection results. Thus, 10 cells per period  represent a sufficiency fine discretization for the chosen angles of incidence and refraction, which results in unit cells with lateral dimensions of $\lambda/9.5\times\lambda/9.5$.} As seen previously in the calculated G parameters, realizing the metasurface essentially becomes equivalent to  designing unit cells which have unity G$_{21}$ magnitude and the appropriate G$_{21}$ phase \cite{Wong2016}. \textcolor{black}{In order to properly match the desired G matrix, three degrees of freedom are needed. Essentially these three degrees of freedom refer to the matching of the incident wave ($|\mathrm{G}_{11}| = 0$), matching of the refracted wave ($|\mathrm{G}_{22}| = 0$), and the desired transmission phase ($\angle \mathrm{G}21$). Therefore, to produce a set of unit cells which can match the desired G parameters, the proposed unit cell will utilize a three-layered structure to match the three degrees of freedom required\cite{Wong2016, Epstein2016_3}.} The unit cells consist of three etched metal layers patterned on two substrates. Each substrate is chosen to be approximately $\lambda/22$ (0.635 mm) thick resulting in a total thickness of approximately $\lambda/11$ (1.27 mm). The overall theoretical structure can be seen in Fig. $\ref{fig:layer}$. As shown, the theoretical structure can be modeled in an equivalent circuit model composed of three shunt admittances separated by transmission lines \cite{Monticone2013,Pfeiffer2014_3, Wong2016, Epstein2016_3}. The shunt impedances will model the metal layers while the transmission lines represent the substrates. By modeling the unit cell in this equivalent circuit, the response of the circuit can be uniquely determined. Essentially, the G matrix of this unit cell, which is a function of the layer shunt admittances Y$_1$, Y$_2$, and Y$_3$, can be calculated. The steps in designing the metasurface unit cells now become straightforward. From the translated G matrix from the boundary conditions, we have obtained the required G$_{21}$ parameters. From our unit cell circuit model, we have determined the relation of its G parameters as a function of the three layer admittances. The obvious next step is then to equate the G parameters of the unit cell to those of the translated boundary conditions, and calculate the required admittance of the metal layers. Applying our refraction scenario, the discretized shunt admittances for the 10 theoretical unit cells are numerically obtained and can be seen in Fig. $\ref{fig:theoryimpede}$.

Now that the shunt admittances of the unit cell layers have been obtained, the next step is to find physical metal geometries which can produce the desired response. However, before proceeding to the physical unit-cell design, a few features of the theoretically calculated structure should be discussed. Examining Fig. $\ref{fig:theoryimpede}$, it is clear that the theoretical unit cell structure is asymmetric. As previously discussed, since these unit cells are used for refraction, the port impedances of these cells are different due to the different angles of the incident and refracted plane waves. In terms of microwave networks, the unit-cell equivalent circuits must be matched to different port impedances while maintaining full transmission of the input power. Obviously the only solution to this would be to have an asymmetric structure \cite{Asadchy2016,Wong2016,Epstein2016_3}. This can actually also be seen in Eq. ($\ref{eq:matrix4}$), where $Z_{11}$ and $Z_{22}$ are different values. However, since this asymmetry arises from the boundary conditions, this translates to the bianisotropy of the metasurface \cite{Wong2016,Epstein2016_3}. 


\subsection{Unit Cell Design}
\label{subsec:design_2}

\begin{figure}
\begin{center}
\vspace{-0.3cm}
\noindent
\includegraphics[scale=0.23]{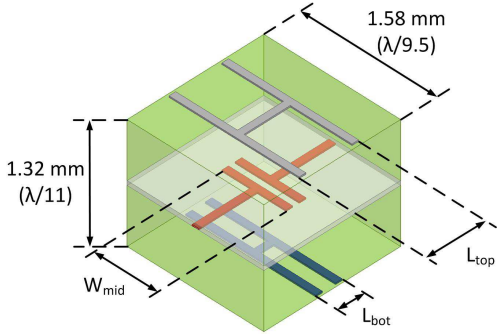}
  \caption{Proposed physical 3-layer unit cell composed of dogbones and a loaded dipole.}
  \label{fig:unitcell}
 \vspace{-0.6cm}
\end{center}
\end{figure}

\begin{figure}
\begin{center}
\vspace{-0.3cm}
\noindent
\includegraphics[scale=0.21]{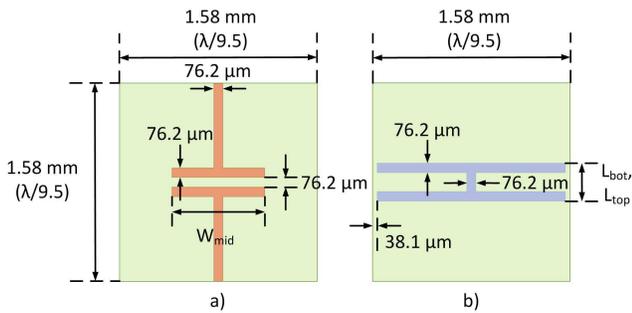}
  \caption{\textcolor{black}{Dimensions of layer geometries used for the proposed 3-layer unit cell. a) Loaded dipole dimensions used for the middle layer. b) Dogbone dimensions used for the top and bottom layers.}}
  \label{fig:unitcell_lat}
 \vspace{-0.6cm}
\end{center}
\end{figure}

To implement the metasurface unit cells, the 3-layer admittance sheet structure of \cite{Wong2016, Epstein2016_3, Epstein2016} was used. The physical unit cells consists of three copper layers (1/2 oz. or  18 $\mu$m thick), etched on two 25 mil (0.635 mm) Rogers RT/duroid 6010 substrates. The two substrates are then bonded together using a 2 mil (0.0508 mm) Rogers 2929 bondply, yielding an overall unit cell thickness of 52 mil (1.3208 mm) which is approximately $\lambda/11$ at the design frequency of 20 GHz. Each unit cell has lateral dimensions of 1.58 mm$\times$1.58 mm or approximately $\lambda/9.5\times\lambda/9.5$, corresponding to 10 unit cells per period as previously mentioned. \textcolor{black}{Each unit cell is formed using a dogbone, a loaded dipole, and another dogbone, on the top, middle, and bottom layers, as shown in Fig. \ref{fig:unitcell} and Fig. \ref{fig:unitcell_lat}.} The response of each layer and thus the overall unit cell is controlled by the L$_\mathrm{bot}$, L$_\mathrm{top}$, and W$_\mathrm{mid}$ of the corresponding layer. To find the appropriate combination of layer geometries for the desired unit cells, the layers are first studied individually. The behavior of each layer as a function of their respective geometric variation is obtained by simulating them individually. 

\textcolor{black}{To extract the behavior of the top layer for example, the middle and bottom metallic layers are removed, but both the dielectric substrates remain. The Floquet ports from HFSS are then referenced to be directly contacting the unit cell. While the ports themselves are placed sufficiently far away to avoid capturing evanescent modes, the phase of the Floquet-port parameters is referenced directly to the physical unit cell. The Floquet-port S parameters are then obtained from HFSS and translated into the equivalent ABCD matrix. This ABCD matrix contains the characteristics of the top metallic layer and the two substrates. To obtain just the effect of the top metallic layer, the substrates can be treated as transmission lines and numerically removed. For the middle layer, a similar approach is taken. In this case, both dielectric substrates remain, and the only metal feature included is the middle layer, with the top and bottom metallic features removed. The Floquet ports again are referenced directly to the unit cell. The ABCD matrix can be once again obtained and the substrates are then similarly numerically removed. The remaining shunt component of the metallic pattern is then characterized.  The performance of the bottom layer can be obtained in the same fashion. As the metallic patterns are simulated and characterized in the presence of the substrates, the effect of the effective dielectric constant seen by the metallic features is characterized. A similar approach can be found in \cite{Pfeiffer2014_3}, where the unit cells are simulated in two half spaces containing air and the dielectric substrates.}

\begin{figure}
\captionsetup[subfigure]{labelformat=empty}
\begin{center}
\vspace{-0.3cm}
\noindent
\subfloat[]{\includegraphics[scale=0.22]{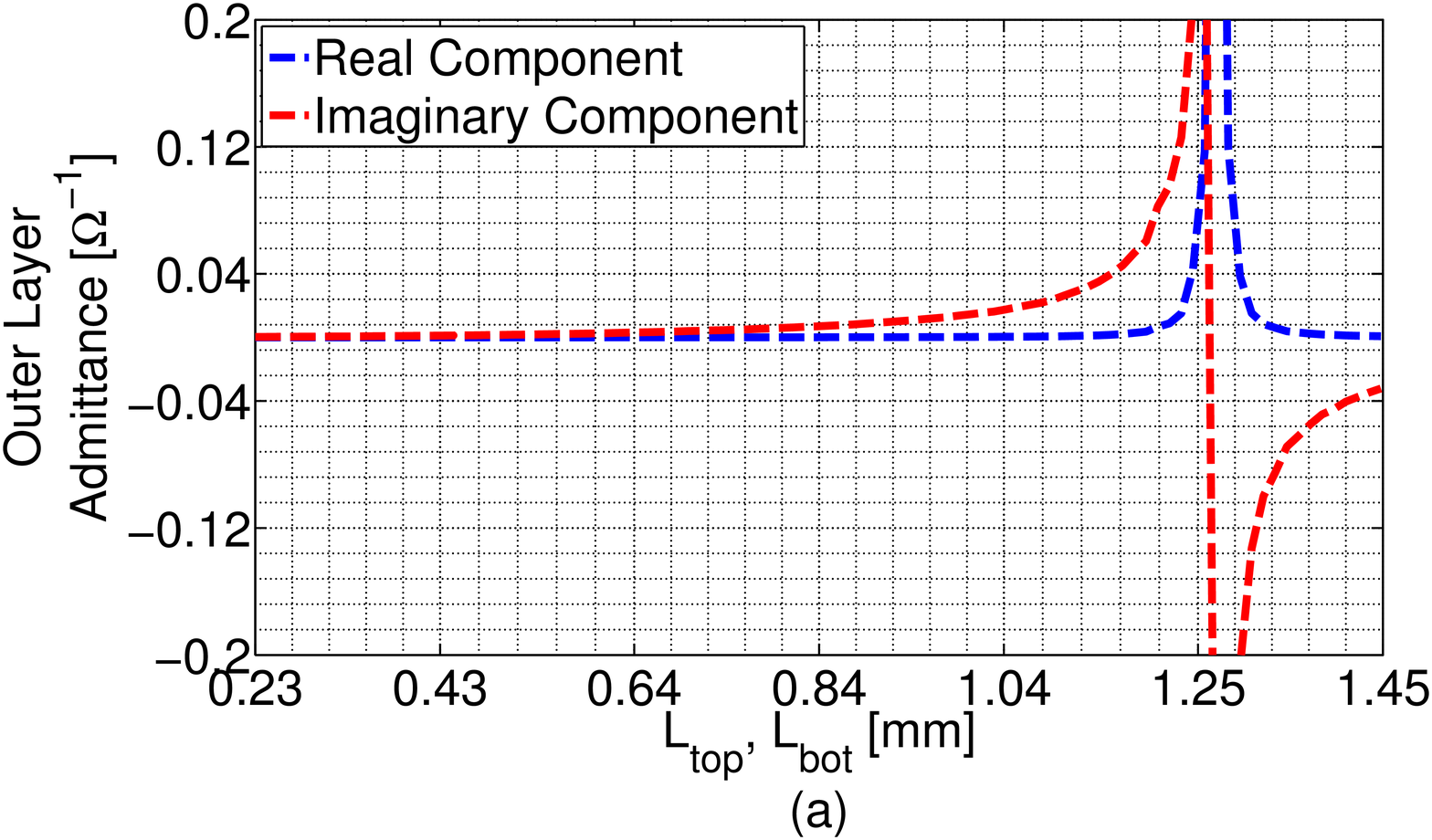}}

\subfloat[]{\includegraphics[scale=0.22]{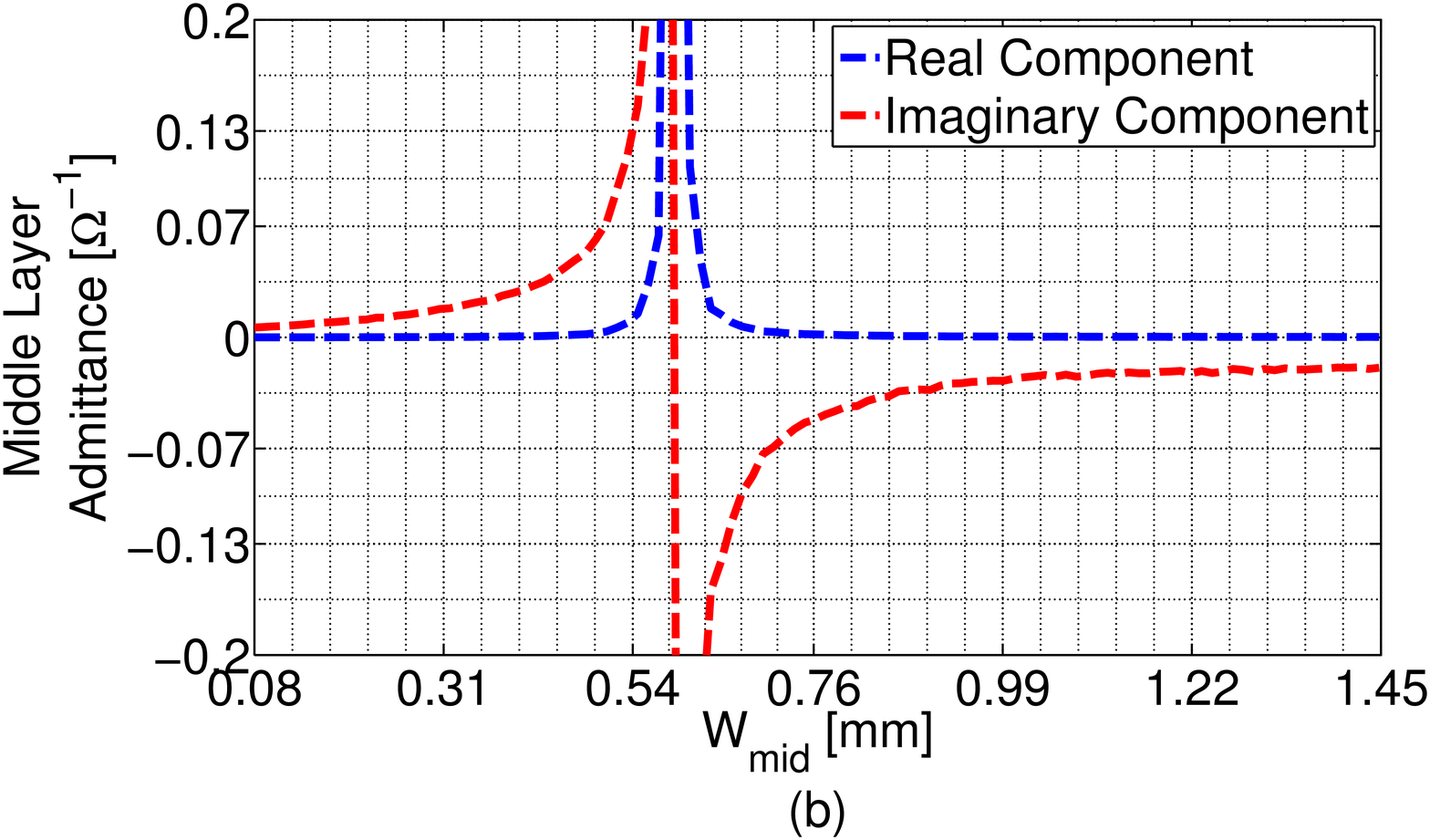}}
  \caption{Admittance response of unit cell layers. a) Outer layer dogbone response as a function of L$_{\mathrm{top}}$, L$_{\mathrm{bot}}$. b) Middle layer loaded dipole response as a function of W$_{\mathrm{mid}}$.}
  \label{fig:dogimpede}
 \vspace{-0.6cm}
\end{center}
\end{figure}

\textcolor{black}{Applying this extraction method, the responses of the middle and outer layers as a function of W$_\mathrm{mid}$, L$_\mathrm{top}$, and L$_\mathrm{bot}$ can be seen in Fig. \ref{fig:dogimpede}.} It is clear that as the layer geometries are varied, their shunt admittances demonstrate both capacitive and inductive responses due to the presence of a geometric resonance point. This is a crucial feature as the theoretically calculated admittances also requires both inductive and capacitive values as previously seen in Fig. \ref{fig:theoryimpede}. Therefore, by utilizing resonant structures, the entire required range of needed shunt admittance values can be obtained. \textcolor{black}{It should be noted that, in general, any scatterer which can implement the desired boundary conditions can be used. In this case, dogbones and a loaded dipole were chosen as they are able to produce both the capacitive and inductive responses as shown in Fig. \ref{fig:dogimpede}. These structures have been demonstrated as capable of producing metasurface designs in the past \cite{Epstein2016_3, Epstein2016}.}

\begin{figure}
\captionsetup[subfigure]{labelformat=empty}
\begin{center}
\vspace{-0.3cm}
\noindent
\subfloat[]{\includegraphics[scale=0.22]{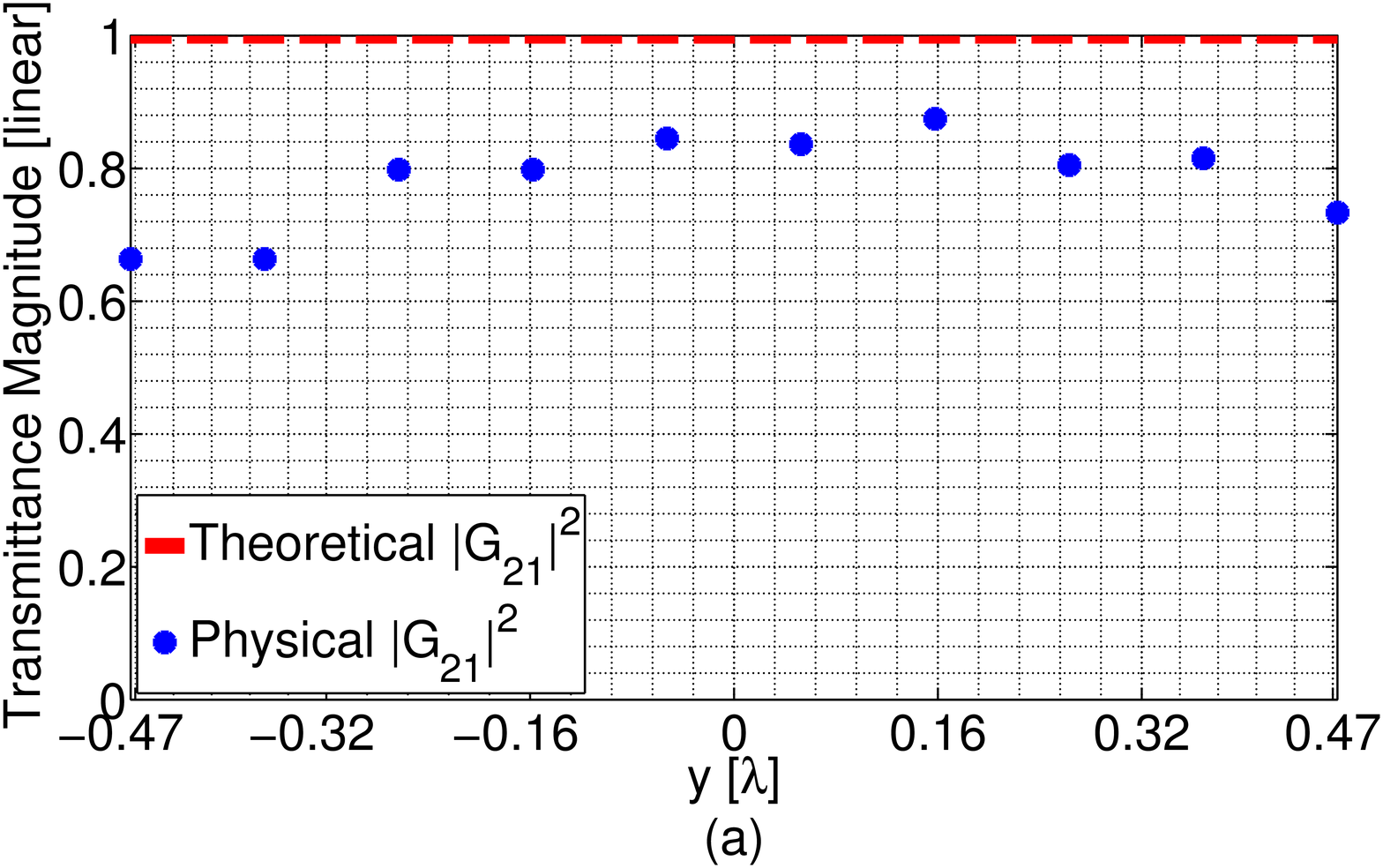}}

\subfloat[]{\includegraphics[scale=0.22]{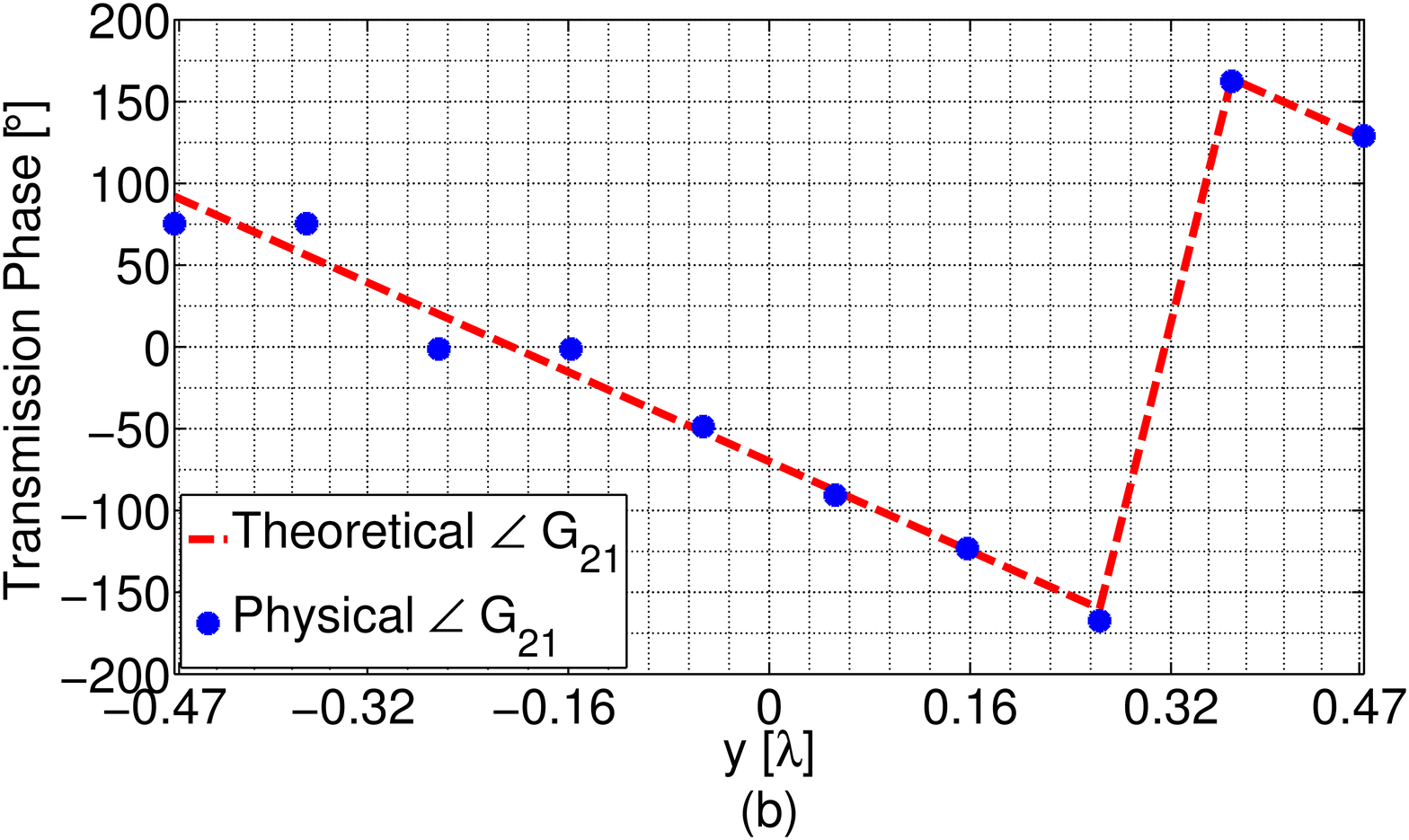}}
  \caption{Comparison of realized and theoretical G parameter response of the designed unit cells. a) Magnitude of G$_{21}$. b) Phase of G$_{21}$.}
  \label{fig:unitcellperform}
 \vspace{-0.6cm}
\end{center}
\end{figure}

Using the above mentioned layer extraction method, the unit cell layer dimensions can be varied to match the desired admittance values shown in Fig. \ref{fig:theoryimpede}. Once the desired admittance of each layer is matched to physical geometries, the full unit cell with all 3 layers can be assembled. The G parameters of the complete unit cells can then be compared to the theoretical G parameters to validate their response. It should be noted that due to coupling of the layers, the finalized unit cell will need tuning to achieve the desired response \cite{Pfeiffer2014_3}. However, the layer by layer matching technique provides an initial design for further optimization. The final element geometrical parameters are presented in Table \ref{tab:unitcell}. Comparison of the physical and theoretical unit-cell responses can be seen in Fig. \ref{fig:unitcellperform}. Although metallic and dielectric losses prevented the G$_{21}$ magnitude to be unity, the overall phase was in good agreement with the theoretical values. It should be noted that due to increased losses exhibited by unit cells with G$_{21}$ phase around 0$^\circ$, we used identical structures for unit-cell pairs \#1 \#2\ and \#3 \#4\ which implemented the average phase response of the two elements. Applying this averaging technique, the losses were reduced and the transmission in simulation was improved.  \textcolor{black}{It should also be noted that unlike \cite{LavigneArxiv2017}, no full-wave optimization of the metasurface period was used to optimize the unit cells.}

\begin{table}
\caption{\label{tab:unitcell}%
Finalized unit cell dimensions for bianisotropic refraction}
\begin{ruledtabular}
\begin{tabular}{ccccc}
Unit Cell Number & L$_{top}$ [mm]& W$_{mid}$ [mm]& L$_{bot}$ [mm]\\
\hline
1 & 0.8382& 0.5334& 0.9144\\
2 & 0.8382& 0.5334& 0.9144\\
3 & 0.5334& 0.8153& 0.1067\\
4 & 0.5334& 0.8153& 0.1067\\
5 & 0.6858& 0.7391& 0.6553\\
6 & 0.7027& 0.7772& 0.7772\\
7 & 0.7143& 0.9449& 0.8077\\
8 & 0.8382& 0.4191& 0.6096\\
9 & 0.8105& 0.5029& 0.7544\\
10 & 0.8001& 0.5334& 0.8382\\
\end{tabular}
\end{ruledtabular}
\end{table}

\subsection{Simulation Results}
\label{subsec:design_3}

\begin{figure}[h]
\begin{center}
\vspace{-0.3cm}
\noindent
\includegraphics[scale=0.3]{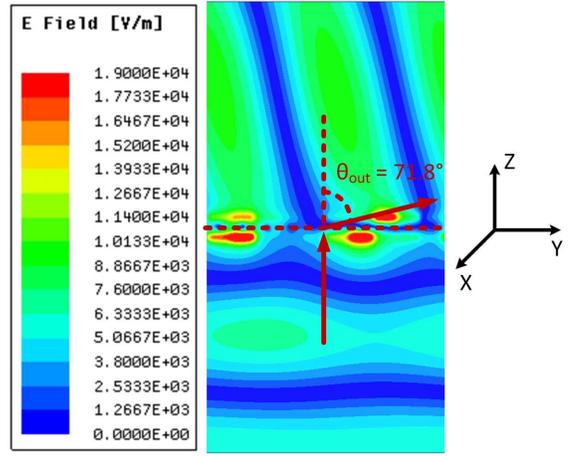}
  \caption{Electric field distribution showing refraction using periodic fullwave simulation}
  \label{fig:HFSSfields}
 \vspace{-0.6cm}
\end{center}
\end{figure}

With the 10 unit cells designed, one period of the metasurface was simulated under periodic boundary conditions in HFSS. The metasurface was then excited by using Floquet ports on the top and bottom of the structure. The Floquet ports are able to extract the transmission and reflections of the associated plane-wave modes which arise from the periodicity of the surface. By investigating the transmission and reflection of each of the Floquet modes, the refraction and reflection nature of the design could be quantified. Additionally, the full field distribution could also be obtained and is shown in Fig. \ref{fig:HFSSfields}.

\begin{figure}
\captionsetup[subfigure]{labelformat=empty}
\begin{center}
\vspace{-0.3cm}
\noindent
\subfloat[]{\includegraphics[scale=0.22]{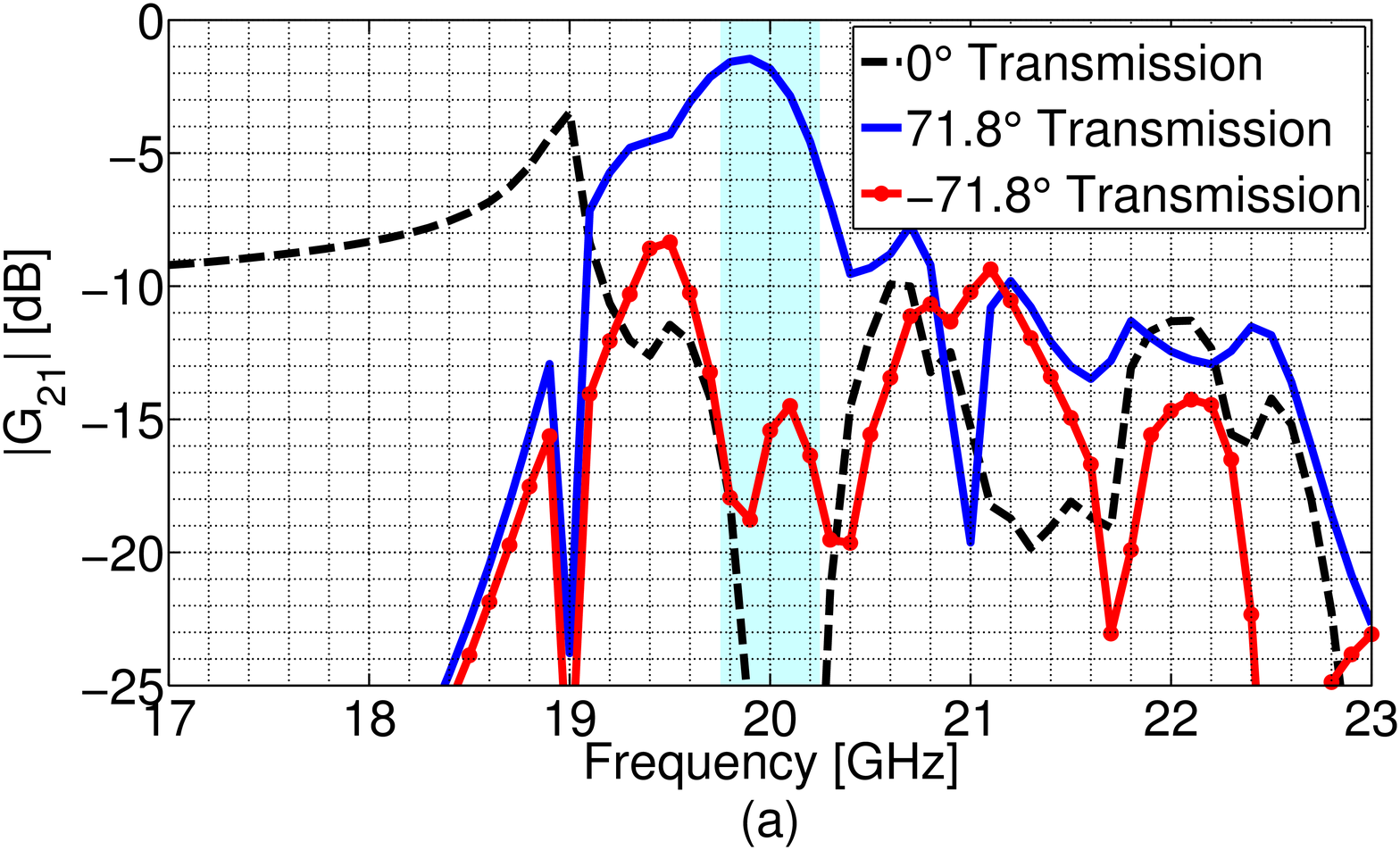}}

\subfloat[]{\includegraphics[scale=0.22]{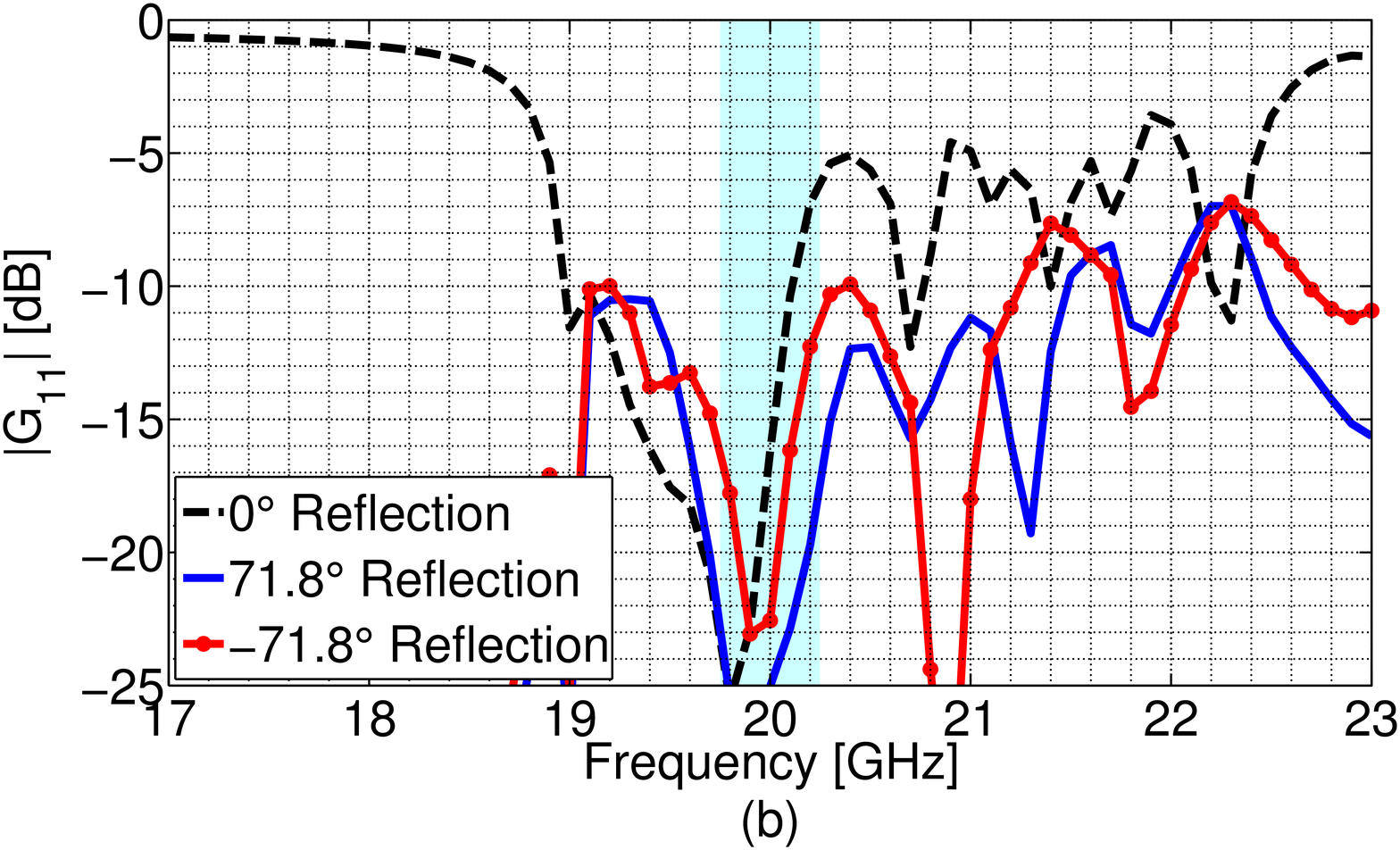}}
  \caption{G parameters from periodic simulation of one period of the metasurface. a) Magnitude of transmission modes, corresponding to the 0$^{\mathrm{th}}$(0$^\circ$), +1(+71.8$^\circ$) and -1(-71.8$^\circ$) transmitted modes. b) Magnitude of reflection modes, corresponding to the 0$^{\mathrm{th}}$(0$^\circ$), +1(+71.8$^\circ$) and -1(-71.8$^\circ$) reflected modes.}
  \label{fig:modes}
 \vspace{-0.3cm}
\end{center}
\end{figure}

Examining the electric field distribution, it is clear that refraction is indeed achieved. On the bottom region of the simulation domain, the metasurface period is excited with a normally incident plane wave. \textcolor{black}{Although some reflection can be noticed from the surface, it does not qualitatively seem significant.} On the top region, the transmitted fields can be seen. It is evident that a clear wave-front refracted at 71.8$^\circ$ is observed.
However, to quantify the characteristics of the design, the simulated G parameters from HFSS are obtained and can be seen in Fig. \ref{fig:modes}.

Due to the macro-periodicity of the boundary conditions, three propagating Floquet modes are allowed to exist. These modes are the 0$^{\mathrm{th}}$ mode, the +1 mode, and the -1 mode, corresponding to plane waves at 0$^\circ$, +71.8$^\circ$ and -71.8$^\circ$ relative to the normal of the metasurface. However, in this case, due to the metasurface possibly generating both reflected and transmitted fields, these three modes could exist in both transmission and reflection regimes. The desired effect of the metasurface would be to have low reflections in all three reflection modes, and having high transmission only in the +1 or +71.8$^\circ$ mode while suppressing the 0$^{\mathrm{th}}$ and -1 transmission modes. Examining Fig. \ref{fig:modes}a, the transmitted modes can be seen. It is clear that around the frequency region of interest, near 20 GHz, the transmission to the +1 or +71.8$^\circ$ mode is maximized while the 0$^{\mathrm{th}}$ and the -1 modes, corresponding to transmission to 0$^\circ$ and -71.8$^\circ$ beams, are strongly suppressed. \textcolor{black}{Additionally, investigating Fig. \ref{fig:modes}b, the reflected modes can be seen. Examining the reflections, it is evident that  all the reflected modes are suppressed at 20 GHz, with reflections all being lower than -15dB.} 

\begin{figure}[h]
\begin{center}
\vspace{-0.3cm}
\noindent
\includegraphics[scale=0.22]{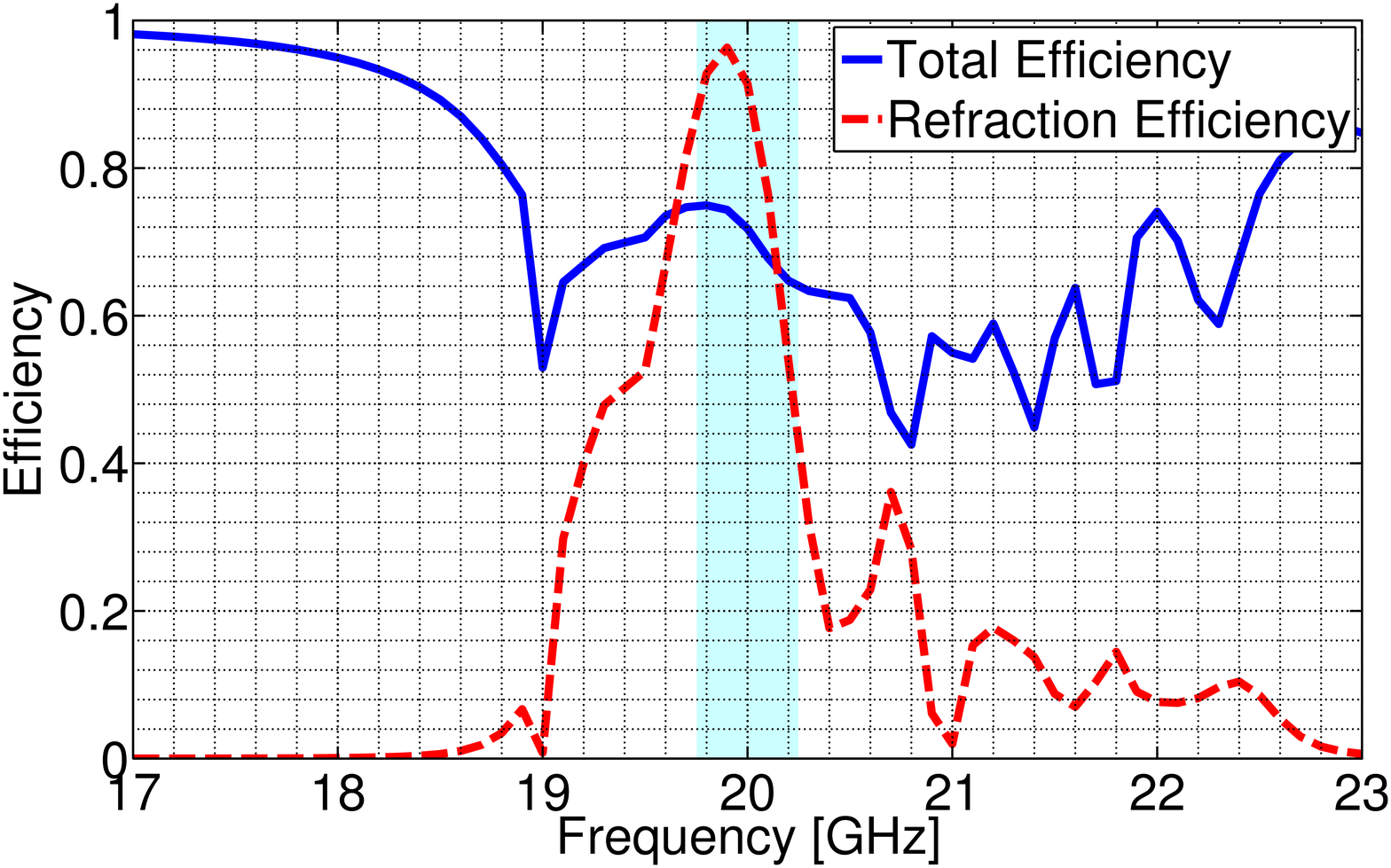}
  \caption{Fullwave simulation of refraction efficiency and total power efficiency.}
  \label{fig:eff}
 \vspace{-0.3cm}
\end{center}
\end{figure}

The metasurface can also be further examined in terms of its total efficiency and refraction efficiency, as shown in Fig. \ref{fig:eff}. The total efficiency is defined as the ratio of the total scattered power to the incident power, while the refraction efficiency as the ratio of the scattered power in the refracted beam to the total scattered power. Investigating the refraction efficiency, $93\%$ of the scattered power is coupled to the desirable Floquet mode (transmitted towards $71.8^\circ$). Combining the high refraction efficiency and low reflections discussed previously, it is clear that the metasurface indeed is able to produce the desired refraction with very high effectiveness.

It should be mentioned that although the refractive and low reflection nature of the metasurface are clearly demonstrated, as seen from the total efficiency, simulations show that approximately $28\%$ of the incident power is absorbed in the metasurface. The relatively high losses probably originate due to the resonant nature of the unit cells. However, ideas on reducing the loss via increasing the number of physical layers have been suggested \cite{Pfeiffer2014_3}. While the design complexity would be increased, designs with additional layers may be considered in the future, providing additional degrees of freedom for improved loss performance. Nonetheless, with the main functionality features of the metasurface, namely the low reflections and high refraction efficiency of the metasurface demonstrated in simulations, a metasurface prototype was fabricated to be experimentally verified.


\section{Experimental Results}
\label{sec:results}

\begin{figure}[h]
\begin{center}
\vspace{-0.3cm}
\noindent
\includegraphics[scale=0.23]{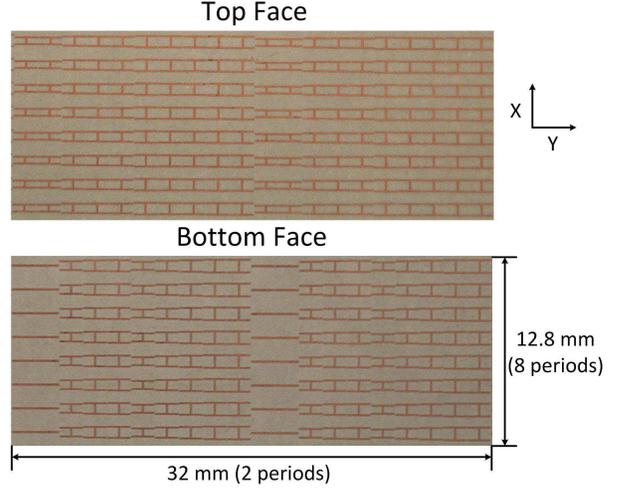}
  \caption{Photo of zoomed in regions of the top and the bottom of the fabricated metasurface.}
  \label{fig:metaphoto}
 \vspace{-0.3cm}
\end{center}
\end{figure}

Following the validation of the metasurface design via fullwave simulations, a metasurface PCB of size 12''$\times$18'' (30.48 cm$\times$45.72 cm), approximately $20\lambda\times30\lambda$, was fabricated. The fabricated surface contains ${29\times190}$ replicas of the simulated metasurface period. A photograph of one section of the metasurface from both sides is shown in Fig. \ref{fig:metaphoto}. To characterize the prototype, two experiments were designed. The first uses a \textcolor{black}{quasi-optical} setup which characterizes the specular reflection properties of the metasurface. The second test utilizes a standard anechoic antenna chamber to test the refraction characteristics of the metasurface. The experiments and results are discussed individually. 

\subsection{\textcolor{black}{Quasi-optical} Specular Reflection Experiment}
\label{subsec:quasi_optical}
\begin{figure}
\centering
\includegraphics[width=7.8cm]{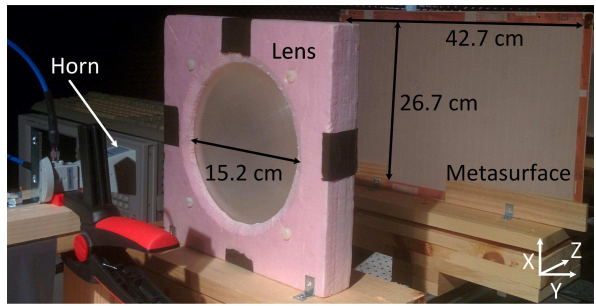}
\caption{\textcolor{black}{Quasi-optical} experimental setup. The focal distances from the lens to the horn and from the lens to the metasurface are 12.5 cm and 29 cm, respectively.}
\label{fig:quasi_optical_setup}
\end{figure}

\begin{figure}[h]
\begin{center}
\vspace{-0.3cm}
\noindent
\hspace{-0.2cm}\includegraphics[scale=0.22]{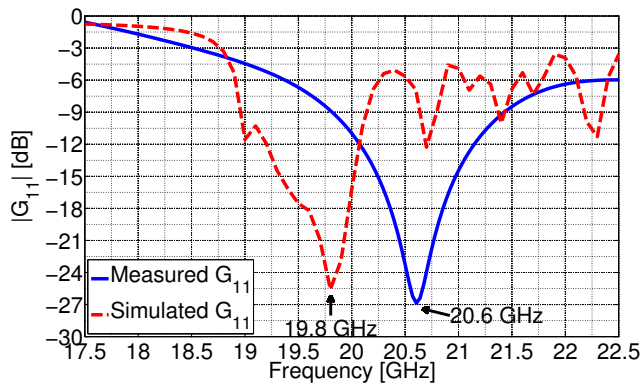}
  \caption{\textcolor{black}{Measured and simulated (HFSS periodic simulation) normally incident specular reflection (0th reflected mode).}}
  \label{fig:S11_response}
 \vspace{-0.6cm}
\end{center}
\end{figure}

To quantify the specular reflectionless nature of the metasurface, a \textcolor{black}{quasi-optical} experiment was designed. The setup is \textcolor{black}{quasi-optical} as it uses a dielectric lens and a horn antenna to establish a line of sight signal link. In this setup we used an A-Info LB-OMT-150220 horn and a bi-convex Rexolite lens to focus a Gaussian beam onto a reference plane. In doing so, the radiation from the horn antenna is collimated by the lens to form a planar wavefront at the reference plane. The horn antenna which is then fed and measured using an Agilent VNA is calibrated to measure the reflection from an object placed at this plane. When no device under test (DUT) is present, the reflections back to the horn antenna are essentially zero. However, when the DUT is inserted, the normally incident reflection can be characterized. Therefore, by using the fabricated metasurface as the DUT, the specular reflection of the 0$^\circ$ incident face of the metasurface could be obtained. The setup with the horn antenna, lens, and metasurface can be seen in Fig. \ref{fig:quasi_optical_setup}. The measured specular reflection of the metasurface is then obtained and is compared to the corresponding simulation results, which can be seen in Fig. \ref{fig:S11_response}. \textcolor{black}{In this case, the simulation results refer to the simulated 0$^{\mathrm{th}}$ order reflection that was shown in Fig. \ref{fig:modes}b. As the \textcolor{black}{quasi-optical} experiment illuminates the metasurface with a planar Gaussian beam, and the beam waist is much smaller than the metasurface, the measurement performed is essentially a plane-wave characterization of the metasurface with minimal edge effects. Thus, the specular reflection of the measured finite metasurface and the simulated infinite surface could be fairly compared.} 


Examining the measured specular reflections, a shift of the resonant frequency to 20.6 GHz can be seen. The source of this shift is most likely attributed to fabrication errors and material parameters. Due to this resonance shift, the experiment results will be focused on 20.6 GHz rather than the nominal 20 GHz. Nonetheless, the measured G$_{11}$ at the resonant frequency indicates that less than 0.2\% of the incident power is back-reflected, which is in agreement with simulations. Additionally, the general trend of the specular reflections match those of the fullwave simulation. Therefore, although there is a frequency shift, at the experimental resonant frequency, the measurements indeed demonstrates the metasurface's near reflectionless specular performance.

\subsection{Far-field Anechoic Chamber Refraction Experiment}
\label{subsec:far_field}

\begin{figure}[h]
\begin{center}
\noindent
\includegraphics[scale=0.1]{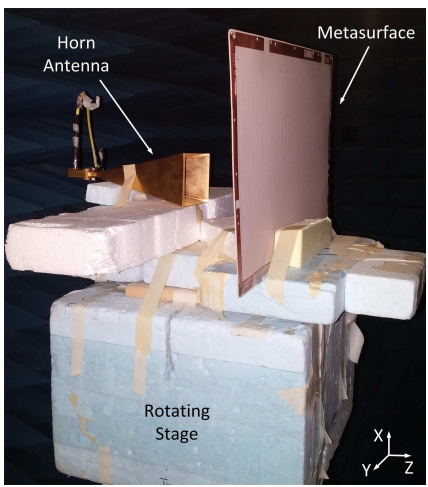}
  \caption{Far-field anechoic radiation measurement setup, with the metasurface and the receiving horn as an overall AUT.}
  \label{fig:FF}
 \vspace{-0.6cm}
\end{center}
\end{figure}

While the \textcolor{black}{quasi-optical} experiment was able to demonstrate the specular reflectionless nature of the metasurface, its refraction properties must also be characterized. To demonstrate if the metasurface is able to refract an incoming wave to the desired angle, an anechoic chamber antenna test was performed. The metasurface was placed in front of a Quinstar QWH-KPRS-00 standard gain horn antenna and the radiation pattern of the overall system, horn with metasurface, was measured. A transmitting horn was placed sufficiently far from the metasurface to produce a planar wavefront, while a receiving horn was aligned behind the metasurface. In this setup, the normally incident side was faced towards the receiving horn and the 71.8$^\circ$ refraction side was facing the transmitting horn, as shown in Fig. \ref{fig:FF}. 
The receiving horn with the metasurface can then be seen as an overall antenna under test (AUT). The AUT was then rotated around its axis, and the gain pattern was measured. To validate the correct refraction effect, the gain pattern of the overall AUT should have maximum gain around the designed angle of 71.8$^\circ$ from broadside. 

While this far-field method was able to characterize the metasurface's refraction characteristics, there are trade-offs in the experimental setup which should be discussed. One such parameter is the distance of the receiving horn from the metasurface. As the metasurface was designed for plane-wave excitation, the receiving horn antenna should be placed sufficiently far away from the metasurface to match a planar wavefront. However due to the finite size of the metasurface, this was not possible. If the distance between the metasurface and horn was too large, the metasurface would not sufficiently shadow the receiving horn antenna. This would undoubtedly create edge diffraction and other measurement errors. On the other hand, if the metasurface to horn distance is too close, the experiment would no longer mimic interactions with plane waves. Therefore, an experimental distance of 24 cm which is roughly 16$\lambda$ was used, which was found to be a good compromise to account for both issues.

\begin{figure}[h]
\begin{center}
\vspace{-0.3cm}
\noindent
\hspace{-0.1cm}\includegraphics[scale=0.22]{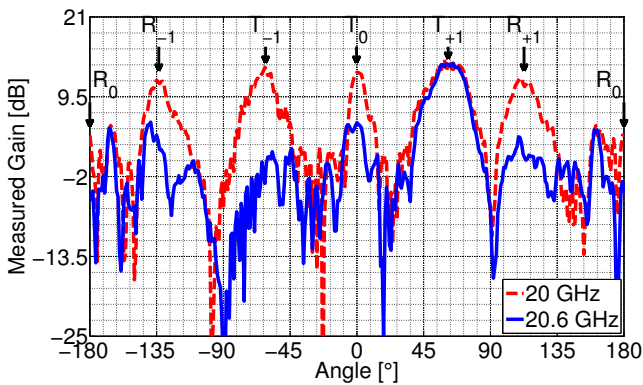}
  \caption{\textcolor{black}{Measured AUT (horn and metasurface) radiation patterns at 20.6 GHz (resonant) and 20 GHz (off resonance). The excited Floquet modes are identified: the 0$^{\mathrm{th}}$ transmitted mode (0$^\circ$), the $\pm1$ transmitted modes ($\pm71.8^\circ$), the 0$^{\mathrm{th}}$ reflected mode ($\pm180^\circ$) and the $\pm1$ reflected modes ($\pm108.2^\circ$).}}
  \label{fig:radiation_pattern}
 \vspace{-0.6cm}
\end{center}
\end{figure}

The measured radiation patterns at 20 GHz and 20.6 GHz can be seen in Fig. \ref{fig:radiation_pattern}. While both frequencies are shown, it should be reminded that due to the frequency shift, 20.6 GHz is the experimental resonance frequency, while the 20 GHz response is used as a comparison for off-resonance performance. From the radiation pattern at 20 GHz (off resonance), all possible propagation modes are present. Additionally, all the modes are measured with relatively equivalent magnitudes. As expected, since the metasurface is tested under off-resonance conditions, the modes which arise due to its macro-periodicity will all be excited. Contrarily, examining the radiation pattern at 20.6 GHz (at resonance), the only Floquet mode that is excited is the desired +1 mode at 71.8$^\circ$. While the +1 mode is unchanged, the other modes in both transmission and reflection regimes are strongly suppressed. 

It should be noted that the only mode that cannot be reliably measured using this far-field measurement is the 0$^{\mathrm{th}}$ or specular reflection mode, due to the blockage of the receiving horn. However, the specular reflection was previously quantified in the \textcolor{black}{quasi-optical} setup, which showed near zero reflections at 20.6 GHz. Thus, combining the \textcolor{black}{quasi-optical} measurements and the far-field measurements, it is evident that the metasurface is indeed able to produce the desired refraction while suppressing all other undesired modes.

\textcolor{black}{The gain pattern at 20.6 GHz also reveals that the peak gain actually occurs at 62$^\circ$. This is due to the effective aperture size of the metasurface. As the metasurface is rotated, the effective aperture size will decrease. This translates into a taper of the measured gain by a cos($\theta$) factor, where $\theta$ is the rotation angle relative to broadside \cite{Chen2017, Skolnik1990}. Therefore, to accurately determine the angle of refraction, this taper must be accounted for. Compensating for this effect, the resulting gain actually peaks at 69$^\circ$, which is close to the desired angle of 71.8$^\circ$. While it is still not a perfect match to the desired angle, the resulting angular deviation can be accounted for due to fabrication tolerances and experimental alignment errors.}


Apart from the refraction angle, the spreading of the refracted beam should also be discussed. In theory, the metasurface is designed as an infinitely periodic structure, however in reality it has a finite size. Additionally as the receiving horn is rather close to the metasurface, the illumination area is even smaller. \textcolor{black}{This is clearly seen in the far-field measurements as the refracted beam has a finite beamwidth due to the finite illumination of the metasurface, whereas in a infinitely periodic setting, the radiation pattern should be a delta function in the direction of refraction.} Using the 3dB beamwidth of the refracted beam at 20.6 GHz, it was calculated that the effective aperture length, in the refraction plane, is roughly 20 cm \cite{Goldsmith1998_Chap7_GaussianHorn}. Given the angular opening and aperture size of the receiving horn, this is a reasonable result. \textcolor{black}{Additionally, as the metasurface prototype has an overall length of 42.7 cm in the refraction plane, the illumination of the horn should produce minimal edge effects in the measurements.}


To characterize the refraction of the metasurface more quantitatively, the measured scattered refraction efficiency is calculated. \textcolor{black}{As previously discussed, as the metasurface prototype has a finite size and illumination, there is an associated beamwidth for the refracted wave. As an effect of the beamwidth, the refracted power is spread out over an angular range. However, as seen from the off-resonance far-field pattern, all the propagating modes experience this effect. Thus, to quantify the scattered refraction efficiency, while accounting for this beam spreading, the integration of the gain of the refracted beam, from null (roughly 30$^\circ$) to null (roughly 90$^\circ$), is compared to the total integration of the 360$^\circ$ gain pattern.} The resulting calculation is the ratio of the refracted power at the desired beam angle over the total scattered power, which demonstrates how efficiently the metasurface is refracting at the designed beam angle. In this case, the scattered refraction efficiency at 20.6 GHz is calculated to be approximately 80\%. While it is lower than the simulated result of 93\%, accounting for experimental errors, finite size of the metasurface, and the resonant frequency shift, the measured 80\% is still a strong indication of efficient refraction. Additionally, as ideal non-bianisotropic \textcolor{black}{Huygens'} metasurface implementing the same wide-angle refraction can only achieve a theoretical efficiency of $73\%$ \cite{Epstein2014_2}, our measured efficiency of 80\% is a good indication of the importance of bianisotropy in metasurface designs.


Fig. \ref{fig:radiation_pattern} also indicates that the suppressed scattering of undesired modes at 20.6 GHz is associated with increased absorption. 
Unfortunately, the current measurement setup does not allow reliable quantification of the losses. One possible source of this loss could be associated with unforeseen material properties, such as anisotropy, and fabrication tolerances. As seen with the resonant frequency shift, the material properties and fabrication uncertainties which produced this shift could have also attributed to higher than expected losses. Another possible source of this increase absorption could be due to the use of resonant structures. As was discussed in Sec. \ref{subsec:design_2}, we used resonant structure for the unit cell design, however due to this resonant nature and the frequency shift, the unit cells may be operating closer to resonance than simulations predicted. Due to this frequency shift, higher losses from the resonant unit cells could also be expected. 
Nonetheless, the losses do not detract the validation of the metasurface as the majority of the scattered power is indeed refracted to the desired mode.



\section{Conclusion}
\label{sec:conclusion}
Metasurfaces have become an indispensable tool for tailoring electromagnetic waves. Applications of metasurfaces such as focusing, refraction and polarization control have been demonstrated in recent years. Refraction, which was originally thought possible by controlling only the electric impedance and magnetic impedance of a metasurface, has proved problematic when the refraction angle was scanned far from normal incidence. \textcolor{black}{The discovery of this issue, brought about the use of bianisotropy via a magnetoelectric coupling coefficient in the boundary condition formulations.} In doing so, wide-angle refraction could be achieved with matched, lossless, and passive metasurfaces. 

In this work, we have demonstrated a PCB bianisotropic metasurface implementing reflectionless wide-angle refraction. The theory and design of the proposed metasurface was presented. The proposed unit cells for demonstrating a 71.8$^\circ$ refraction metasurface for normally incident planes waves at 20 GHz were discussed. Simulations of both the unit cells and a period of the metasurface under periodic conditions were conducted via HFSS fullwave simulations. The results showed promising performance and a PCB prototype was fabricated. Finally, experimental validation of the prototype was carried out by combining the results of a \textcolor{black}{quasi-optical} setup and radiation pattern measurements. Through the \textcolor{black}{quasi-optical} experiment, the specular reflections of the metasurface were demonstrated to be minimal. By utilizing the far-field radiation pattern measurement, the refraction of the metasurface was validated. This hybrid approach verifies that, indeed, the specular reflections of the metasurface are negligible, and approximately $80\%$ of the scattered power is coupled to the desired beam. Although a resonant frequency shift may have resulted in increased losses, the overall refraction properties of the metasurface were validated. 


\begin{acknowledgments}
\textcolor{black}{
Financial support from the Natural Sciences and Engineering
Research Council of Canada (NSERC) is gratefully acknowledged.}

\textcolor{black}{
This project has received funding from the European Union's Horizon 2020 
research and innovation program under the Marie Sklodowska-Curie grant 
agreement No 706334.} 
\end{acknowledgments}

\bibliography{refraction_bib}

\end{document}